# Memristive Model of Excitable Cells

Maheshwar Sah and Ram Kaji Budhathoki

**This paper presents in-depth analysis of the excitable membranes of a biological system. We rigorously prove from the Chay neuron model that the state dependent voltage-sensitive potassium ion-channel and calcium sensitive potassium ion-channel in excitable cells are in-fact generic memristors and state independent mixed sodium and calcium ion-channel is non-memristive (nonlinear resistor) element in the perspective of electrical circuit theory. The mechanism to give the rise of the periodic oscillation, aperiodic (chaotic) oscillation, spikes and bursting in excitable cells are also analyzed via the small-signal model, pole-zero diagram, local-activity principle, edge of chaos and Hopf-bifurcation theorem. It is also shown that the presence of complex-conjugate and positive real part of zeros (equivalent to the Eigen values) of the admittance function inside the two bifurcation points lead to the generation of complicated electrical signals in excitable membrane.**

*Keywords:* **Memristor; excitable cells; oscillation; chaos; spikes; bursting; Chay model; small-signal model; pole-zero diagram; local activity; edge of chaos; Hopf bifurcation**

## 1. Introduction

The generation of voltage oscillation, action potential, spikes, chaos and bursting in biological membranes have been studied, investigated and observed experimentally by many researchers over a century. The popular mathematical and electrical circuit model developed by Hodgkin-Huxley(HH) in 1952 [Hodgkin & Huxley, 1952] consisting the membrane voltage, potassium conductance, sodium conductance and leakage conductance describes the propagation of action potential based on the experimental data of squid giant axon. It was identified that the potassium ion and the sodium ion in the HH model misidentified as a time-varying potassium conductance and a time-varying sodium conductance are in fact generic memristors respectively from the perspective of electrical circuit theory [Chua & Kang, 1976, Chua et al., 2012a, 2012b; Sah et al., 2014, Chua, 2015]. The HH model attracted enormous interests to design a model and observe the experimental results in the wide varieties of complex system of the membrane potential, nervous system, barnacle giant muscle fibre, Purkinje fibers, solitary hair cells, auditory periphery and so on[Hodgkin & Keynes 1956; Morris & Lecar, 1981; Noble, 1962; Hudspeth & Lewis, 1988; Giguère & Woodland, 1994]. Similarly, extensive researches have been conducted to observe the varieties of oscillations in β-cells of the pancreas inspired by the HH model. The model of excitable membrane in pancreatic β–cells [Plant, 1981; Chay 1983; Chay & Keizer 1983] consist voltage-sensitive channels that allow the $Na^+$ and $Ca^{2+}$ to enter the cell and, voltage-sensitive $K^+$ channels and voltage-insensitive $K^+$ channel which allow to leave $K^+$ ion and activate intracellular calcium ion respectively. Therefore, the outward current carried by $K^+$ ions passes through the voltage and calcium-sensitive channels, and inward current carried by $Na^+$ and $Ca^{2+}$ passes through the voltage-sensitive $Na^+$ and $Ca^{2+}$ channels. However, the above models consist of complicated nonlinear differential equations associated with membrane voltage. Later a modified model was presented by Chay [Chay, 1985], assuming the β-cells of the voltage-sensitive $Na^+$ conductance is almost inactive, and the inward current is almost exclusively carried by $Ca^{2+}$ ions through the voltage-sensitive $Ca^{2+}$ channel. Therefore, the assumption of a *mixed* effective conductance was formulated without affecting the results by expressing the total inward current in terms of a single mixed conductance $g_I$, and reversal potential $E_I$ of the two functionally independent $Na^+$ and $Ca^{2+}$ channels. The model consists of three nonlinear differential equations in contrast to the other complicated model of the excitable membrane of pancreatic β-cells. In this paper, our study is focused in the simplified Chay model [Chay, 1985].

Fig. 1(a) shows the typical circuit of Chay model with external current stimulus $I^1$. It consists membrane potential $V$ of capacitance $C_m$, potentials $E_I$, $E_K$ and $E_L$ for mixed $Na^+$-$Ca^{2+}$ ions, $K^+$ and leakage ions respectively, and $g_I$, $g_{K,V}$, $g_{K,Ca}$, and $g_L$, are the conductance of the *voltage-sensitive mixed ion-channel*, *voltage-sensitive potassium ion-channel*, *calcium-sensitive potassium ion-channel* and *leakage channel* respectively. The equivalent memristive model of $g_{K,V}$, $g_{K,Ca}$

Corresponding Author: Maheshwar Sah(sahmaheshwar@gmail.com)

---

[1] Electrical model is not given in the original Chay paper [Chay, 1985]. We have designed the typical circuit following the differential equation of the membrane potential. The symbolic representation of the conductance and potentials which are assumed slightly in different notations compared to the original representation. Fig. 1(a) has shown following the conventional assumption of HH model. The external stimulus $I$ is assumed zero throughout this study.

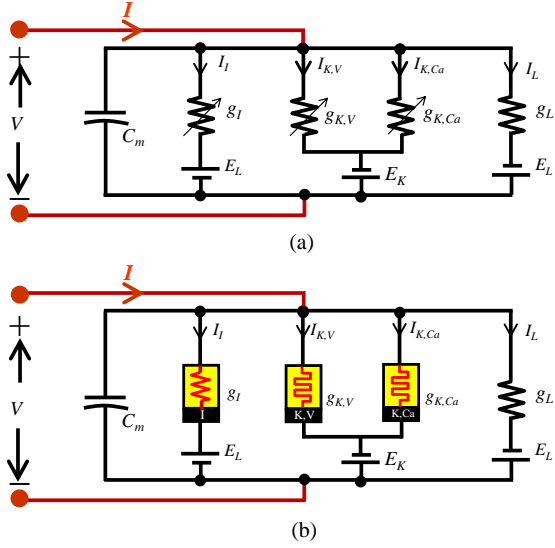

(a)

(b)

**Fig. 1.** Typical electrical circuit of Chay model [Chay, 1985]. (a) Electrical circuit following conventional assumption of HH model [Hodgkin & Huxley, 1952]. (b) Equivalent memristive Chay model. The potential $E_{Ca}$ for $Ca^{2+}$ ion given in the rate of the calcium concentration in (1c) is not shown in Fig. 1(a) and Fig. 1(b).

which will be proven in the next section has been shown in Fig. 1(b). The model defined by three nonlinear differential equations are given in Table 1 and parameters for this model are summarized in Table 2[2].

The functionality of excitable membranes in biological systems are complicated and mechanism to generate periodic, aperiodic (chaotic), bursting and spikes signals in the cells are still under investigation. The aim of this paper is to verify that state in-dependent *voltage-sensitive mixed ion channel $g_I$* is nonlinear resistor and state dependent *voltage-sensitive potassium channel $g_{K,V}$, calcium-sensitive potassium channel $g_{K,Ca}$* are time-invariant memristors. Another goal of this paper is to analyze the mechanism of the generation of complicated electrical signals in excitable cells via small signal equivalent circuit model, pole-zero diagram, local activity principle, edge of chaos and Hopf bifurcation theorem.

## 2. Ion-channel Memristor in terms of Generic memristor

A generic memristor driven by a current source or voltage source is a two-terminal electrical circuit whose instantaneous current or voltage obeys a state-dependent ohm's law. A generic memristor driven by a current source can be expressed as follows in terms of state $\dot{x}_n$:

$$v = R(x_1, x_2, ..., x_n)i \quad (2a)$$

$$\dot{x}_n = f_1(x_1, x_2, ..., x_n; i) \quad (2b)$$

where $R(x)$, depends on "n" ($n \geq 1$) states variables $x = x_1, x_2, ..., x_n$, is the memristance of the memristor.

**Table 1.** Chay model Equations

$$\frac{dV}{dt} = \frac{I - g_I m_\infty^3 h_\infty (V - E_I) - g_{K,V} n^4 (V - E_K) - g_{K,Ca} \frac{Ca}{1+Ca}(V - E_K) - g_L (V - E_L)}{C_m} \quad (1a)$$

$$\frac{dn}{dt} = \frac{n_\infty - n}{\tau_n} \quad (1b)$$

$$\frac{dCa}{dt} = -\rho \left[ m_\infty^3 h_\infty (V - E_{Ca}) + k_{Ca} Ca \right] \quad (1c)$$

where
$n_\infty = \dfrac{\alpha_n}{\alpha_n + \beta_n}$ $\quad \alpha_n = \dfrac{0.01(V+20)}{1 - e^{-0.1(V+20)}}$ $\quad \beta_n = 0.125 e^{\left(\frac{-(V+30)}{80}\right)}$

$m_\infty = \dfrac{\alpha_m}{\alpha_m + \beta_m}$ $\quad \alpha_m = \dfrac{0.1(V+25)}{1 - e^{-0.1(V+25)}}$ $\quad \beta_m = 4 e^{\left(\frac{-(V+50)}{18}\right)}$

$h_\infty = \dfrac{\alpha_h}{\alpha_h + \beta_h}$ $\quad \alpha_h = 0.07 e^{\left(\frac{-(V+50)}{20}\right)}$ $\quad \beta_h = \dfrac{1}{1 + e^{-0.1(V+20)}}$ $\quad \tau_n = \dfrac{1}{\lambda_n (\alpha_n + \beta_n)}$

---

[2] The conductance of calcium sensitive potassium channel memristor $g_{KCa}$ is not listed in Table 2 as it is input parameter throughout this study.

**Table 2.** Parameters values used for the Chay model

| | | | |
|---|---|---|---|
| $C_m$ | $1\mu m/cm^2$ | $g_{K,V}$ | $1700\ s^{-1}$ |
| $E_K$ | $-75\ mV$ | $g_I$ | $1800\ s^{-1}$ |
| $E_I$ | $100\ mV$ | $g_L$ | $7\ s^{-1}$ |
| $E_L$ | $-40\ mV$ | $g_{K,Ca}$ | - |
| $E_{Ca}$ | $100\ mV$ | $K_{ca}$ | $3.3/18\ mV$ |
| $\lambda_n$ | $230$ | $\rho$ | $0.27\ mV^{-1}s^{-1}$ |

Similarly, a voltage-controlled memristor is defined in terms of the memductance $G(x)$ and the state variables $x_1, x_2..., x_n$, as follows:

$$i = G(x_1, x_2, ..., x_n)v \quad (3a)$$

$$\dot{x}_n = f_1(x_1, x_2, ..., x_n; v) \quad (3b)$$

Eqs. (2) and (3) are the core equations to distinguish the memristive and non-memristive system and are used to prove the *voltage-sensitive potassium ion-channel* and *calcium-sensitive potassium ion-channel* are in fact time-invariant generic memristors and *voltage-sensitive mixed ion channel* is non-memrisitve (nonlinear resistor) element.

### 2.1 Voltage-sensitive potassium ion-channel memristor

Let us define the voltage across the *voltage-sensitive potassium ion-channel* shown in third (from left) element in Fig. 1(a) is $v_{K,V}$ and current is $i_{K,V}$, then

$$V - E_K = v_{K,V} \quad (4a)$$

and current entering to the channel is

$$i_{K,V} = G_{K,v}(n) v_{K,V} \quad (4b)$$

where the memductance is given by

$$G_{K,V}(n) = g_{K,V}\ n^4 \quad (4c)$$

and the state equation describing the channel in terms $n$ can be simplified from 1(b) as,

$$\frac{dn}{dt} = f(n; v_{K,V}) = \lambda_n \left[ \frac{0.01(v_{K,V} + E_K + 20)}{1 - e^{-0.1(v_{K,V} + E_K + 20)}}(1-n) - 0.125 e^{\left(\frac{-(v_{K,V}+E_K+30)}{80}\right)} n \right] \quad (4d)$$

Note that (4b)-(4d) are identical to the voltage controlled generic memristor defined in (3a)-(3b) with first order differential equation. Hence, the time-varying conductance shown in Fig. 1(a) of *voltage-sensitive potassium ion-channel* is replaced with *voltage-sensitive potassium ion-channel* memristor as shown in the third element (from left) in Fig. 1(b).

We observed the memristive fingerprint of the

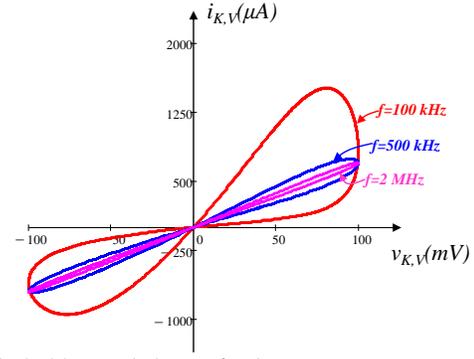

**Fig. 2.** Pinched hysteresis loops of *voltage-sensitive potassium ion-channel* memristor at frequencies $f=100$ KHz, $500$ KHz, and $2$ MHz for the input signal $v_{K,V}(t) = 100\sin(2\pi f t)$.

*voltage-sensitive potassium ion-channel* memristor by applying sinusoidal bipolar signal under different frequencies. This property asserts that beyond some frequency $f^*$, the pinched hysteresis loops characterized by a memristor shrinks to a single-valued function through the origin as frequency $f > f^*$ tends to infinity. To verify this property, a sinusoidal voltage source $v_{K,V}(t)=100\sin(2\pi f t)$ is applied with frequencies $f=100$ KHz, $500$ KHz, and $2$ MHz respectively. As shown in Fig. 2, the zero crossing pinched hysteresis loops shrink as the frequencies increase and tend to a straight line at $2$ MHz which confirms that the *voltage-sensitive* potassium ion-channel is a generic memristor [Chua 2014]. All of these pinched hysteresis loops exhibit the fingerprints of a memristor [Adhikari *et al.,* 2013].

### 2.2 Calcium-sensitive potassium ion-channel Memristor

Let us consider the input voltage of the *calcium-sensitive potassium ion-channel*, the fourth element (from left) in Fig. 1(a) is is $v_{K,Ca}$[3] and current is $i_{K,Ca}$ then the current entering to the channel is given by

$$i_{K,Ca} = G_{K,Ca}(Ca) v_{K,Ca} \quad (5a)$$

where

$$V - E_K = v_{K,Ca} \quad (5b)$$

and the memductance of the calcium-sensitive potassium channel is given by

$$G_{K,Ca}(Ca) = g_{K,Ca}\frac{Ca}{1 + Ca} \quad (5c)$$

The state equation in terms of calcium concentration from (1c) is given by

$$\frac{dCa}{dt} = f(Ca; V_{K,Ca}) = -\rho \left[ m_\infty^3 h_\infty \left(v_{K,Ca} + E_K - E_{Ca}\right) + k_{Ca} Ca \right] \quad (5d)$$

---

[3] Since the same potential $E_K$ is shared by the *voltage-sensitive potassium ion-channel* memristor and *calcium- sensitive potassium ion-channel* memristor, the voltage assumed $V-E_K = v_{K,V}$ in (4a) and $V-E_K = v_{K,Ca}$ in (5b) are basically same. The voltages $v_{K,V}$ and $v_{K,Ca}$ are assumed to distinguish the input voltage applied to *voltage-sensitive potassium ion-channel* memristor and *calcium- sensitive potassium ion-channel* memristor, respectively.

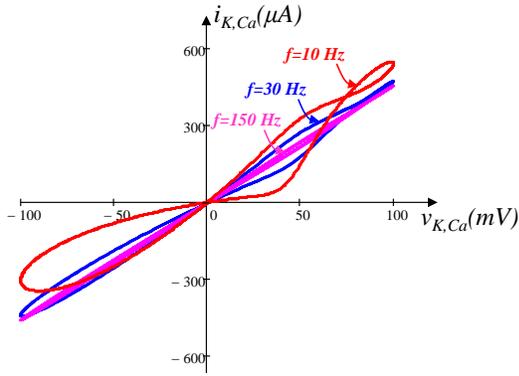

**Fig. 3.** Pinched hysteresis loops of calcium-sensitive potassium channel memristor at frequencies $f = 10Hz, 30Hz$ and $150Hz$ for the input signal $v_{K,Ca}(t) = 100\sin(2\pi ft)$ when $g_{KCa}=10\ s^{-1}$.

Observe that (5b)–(5d) are an example of a voltage-controlled memristor defined in (3a)–(3b) in terms of the calcium concentration channel Ca. Since only one state equation is defined in terms of Ca, we call this memristor as a first order *calcium-sensitive potassium ion-channel* generic memristor. Therefore the time varying *calcium-sensitive potassium ion-channel* is replaced with *calcium-sensitive potassium ion-channel* memristor as shown in the fourth element (from left) in Fig. 1(b).

Let us verify the fingerprint of the frequency-dependent pinched hysteresis loop of the *calcium-sensitive potassium channel* by applying sinusoidal voltage source $v_{KCa}(t) = 100\sin(2\pi ft)$ with frequencies $f=10\ Hz,\ 30\ Hz$ and $150\ Hz$ respectively. Observe from Fig. 3 that, all the zero crossing pinched hysteresis loops shrink as the frequencies of the input signal increase and tend to a straight line for the frequency $f=150\ Hz$. All of the pinched hysteresis fingerprint confirm that the *calcium-sensitive potassium* channel is a generic memristor.

### 2.3 Voltage-sensitive mixed ion channel nonlinear resistor

The time varying *voltage sensitive mixed ion-channel* with input voltage $v_I$ and current $i_I$ in the second element (from left) in Fig. 1(a) is given by,

$$V - E_I = v_I \tag{6a}$$

where

$$i_I = G_I(m,h)\,v_I \tag{6b}$$

and the memductance of the voltage sensitive mixed ion channel is given by

$$G_I(m,h) = g_I m_\infty^3 h_\infty \tag{6c}$$

where

$$m_\infty = \frac{0.1(v_I + E_I + 25)}{0.1(v_I + E_I + 25) + 4\left(1 - e^{-0.1(v_I + E_I + 25)}\right)e^{\left(\frac{-(v_I + E_I + 50)}{18}\right)}} \tag{6d}$$

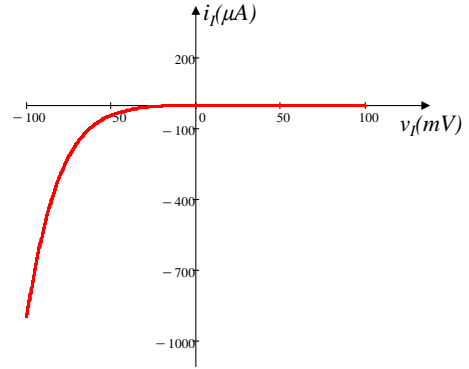

**Fig. 4.** Output waveform plotted in $i_I$ vs $v_I$ plane when input voltage $v_I = 100\sin(2\pi ft)$, with frequencies $f = 100\ Hz,\ 200\ Hz,\ 1\ KHz,$ applied to the *voltage-sensitive mixed ion-channel*. The nonlinear curve shown in Fig. 4 confirms that *mixed ion-channel* is a non-memristive element.

$$h_\infty = \frac{0.07\left(1 + e^{-0.1(v_I + E_I + 20)}\right)e^{\left(\frac{-(v_I + E_I + 50)}{20}\right)}}{0.07\left(1 + e^{-0.1(v_I + E_I + 20)}\right)e^{\left(\frac{-(v_I + E_I + 50)}{20}\right)} + 1} \tag{6e}$$

Observe (6b)–(6e) are not identical to (2a)-(2b) or (3a)–(3b) in terms of state dependent ohm's law. Therefore, the time varying *voltage-sensitive mixed ion-channel* can be replaced by a nonlinear resistor as shown in the second element (from left) in Fig. 1(b). To verify the *voltage-sensitive mixed ion-channel* is not a memristor, a sinusoidal voltage source $v_I = 100\sin(2\pi ft)$ is applied with frequencies $f=100\ Hz,\ 200\ Hz,\ 1\ KHz$ and so on. Fig. 4 shows the corresponding nonlinear waveform of the input voltage $v_I$ vs. output current $i_I$ for any input frequencies. Observe from Fig. 4 that only nonlinear curve is obtained which confirms the *voltage-sensitive mixed ion channel is a non-memristive element* (nonlinear resistor).

### 3. Memristive DC Chay Model

The memristive DC Chay model is obtained by equating the state of the membrance voltage $V$, gate activation $n$ of the *voltage-sensitive potassium ion-channel* memristor and concentration of calcium-sensitive $Ca$ of the *calcium sensitive potassium ion channel* memristor to zero from (1a), (1b) and (1c) respectively and solving for the DC equilibrium point as function of current $I$, i.e.

$$n = n_\infty(V) \triangleq \hat{n}(V) \tag{7a}$$

$$Ca = Ca_\infty(V) \triangleq C\hat{a}(V) \tag{7b}$$

$$I = g_I m_\infty^3 h_\infty(V - E_I) + g_{K,V}\hat{n}^4(V - E_K) + g_{K,Ca}\frac{C\hat{a}}{1+C\hat{a}}(V - E_K) + g_L(V - E_L) \tag{7c}$$

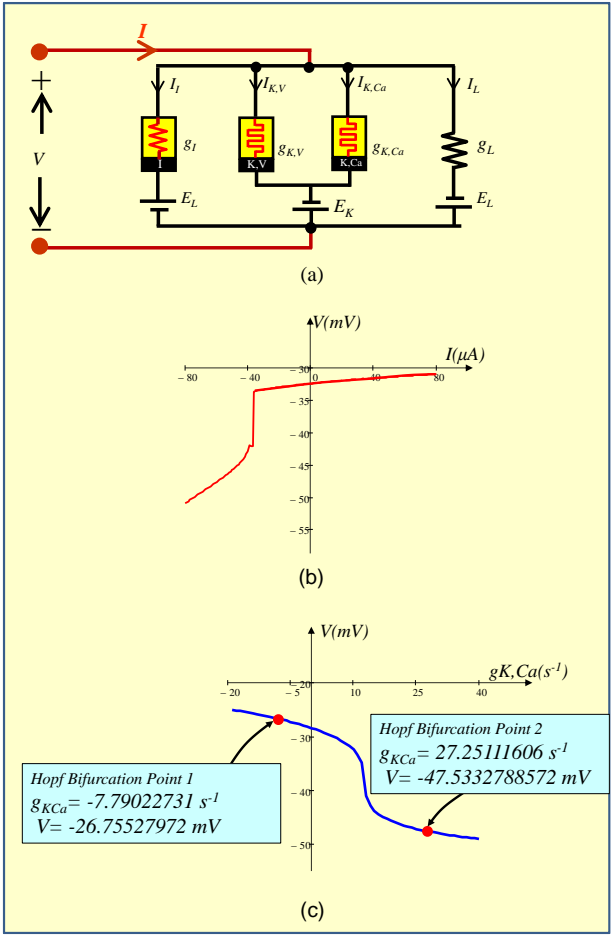

**Fig. 5.** (a) Memristive DC Chay Model at equilibrium voltage $V=V_Q$. (b) DC V-I curve over the range $-80\ \mu A \leq I \leq 80\ \mu A$ at $g_{K,Ca}=10\ s^{-1}$. (c) Membrane voltage $V$ vs $g_{KCa}$ curve over the range $-20\ s^{-1} \leq g_{K,Ca} \leq 40\ s^{-1}$ when $I=0$.

The external current $I$ expressed as the function of membrane voltage $V$ in (7c) gives the *explicit* formula of the DC V-I curve of the memristive Chay model and is shown in Fig. 5(a)[4]. The DC V-I curve for the input current $-80\ \mu A \leq I \leq 80\ \mu A$ when $g_{K,Ca}=10\ s^{-1}$ is shown in Fig. 5(b). Fig. 5(c) shows $V$ vs $g_{K,Ca}$ curve over the range $-20\ s^{-1} < g_{K,Ca} < 40\ s^{-1}$ when external stimulus $I=0$. Our extensive calculations show that, the two Hopf bifurcations points occur at $g_{kCa}= -7.79022731\ s^{-1}$ (resp., $V= -26.75527972\ mV$) and $g_{kca}= 27.25111606\ s^{-1}$ (resp., $V=-47.5332788572\ mV$). Details of these two bifurcation points will be discussed in following section.

## 4. Small-Signal Circuit Model

The small-signal equivalent circuit is the linearized method to predict the response of electronic circuits when a small input signal is applied to an equilibrium point $Q$. The objective of this section is to analyze the small-signal response of

---

[4] The DC V-I curve shown in Fig. 5(b) at $g_{K,Ca}=10\ s^{-1}$ over the range $-80\ \mu A \leq I \leq 80\ \mu A$ is just for simulation purpose. The external current $I$ is assumed to be always zero throughout this study.

*voltage-sensitive mixed ion channel* nonlinear resistor, *voltage-sensitive potassium ion channel* memristor and *calcium-sensitive potassium ion channel* memristor using Taylor series and Laplace transformation.

### 4.1 Small-signal circuit model of the mixed ion-channel nonlinear resistor

The small signal equivalent circuit of the *mixed ion-channel* nonlinear resistor at an equilibrium point $Q_I$[5] on the DC $V_I$-$I_I$ curve is derived as follows

$$v_I = V_I(Q_I) + \partial v_I \tag{8a}$$

$$i_I = I_I(Q_I) + \partial i_I \tag{8b}$$

Applying Taylor series expansion to the *voltage-sensitive mixed ion-channel* nonlinear resistor defined in (8a)-(8b) at the DC operating point $Q_I$, we get

$$\begin{aligned} i_I &= f(v_I + \delta v_I) = a_{00}(Q_I) + a_{12}(Q_I)\delta v_I + h.o.t. \\ &= I_I(Q_I) + \delta i_I \end{aligned} \tag{8c}$$

where,

$$a_{00}(Q_I) = G_I(Q_I)V_I(Q_I) = I_I(Q_I) \tag{8d}$$

$$a_{12}(Q_I) = \frac{\partial f(v_I)}{\partial v_I} \tag{8e}$$

Linearize (8c) by neglecting the h.o.t. then,

$$\boxed{\delta i_I = a_{12}(Q_I)\delta v_I} \tag{8f}$$

Taking the Laplace transform of (8f), we obtain

$$\hat{i}_I(s) = a_{12}(Q_I)\hat{v}_I(s) \tag{8g}$$

The admittance $Y_I(s; Q_I)$ of the small-signal equivalent circuit of *the voltage sensitive mixed ion-channel* nonlinear

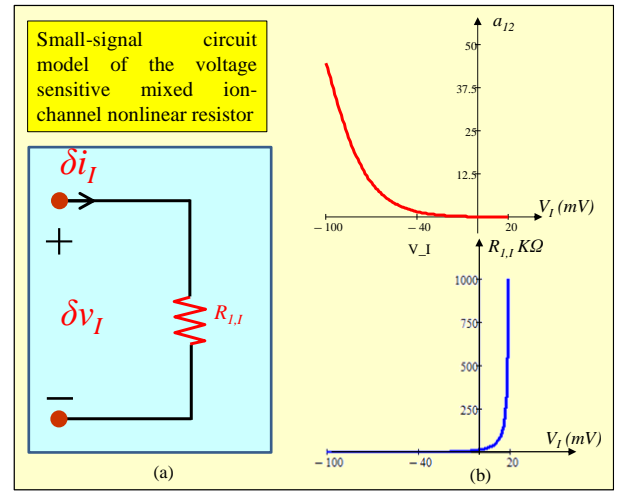

**Fig. 6.** (a) Small-signal circuit model of the *voltage-sensitive mixed ion-channel* nonlinear resistor about the DC equilibrium point $Q_I$ ($V_I$, $I_I$). (b) Plot of the coefficient $a_{12}$ and resistance $R_{1,I}$ as a function of the DC equilibrium voltage $V_I$.

---

[5] The equilibrium point $Q_I$ at $v_i=V_I$ is obtained by solving 6(b).

**Table 3.** Explicit formulas for computing the coefficients $a_{12}(Q_I)$ of the *voltage-sensitive mixed ion channel* nonlinear resistor

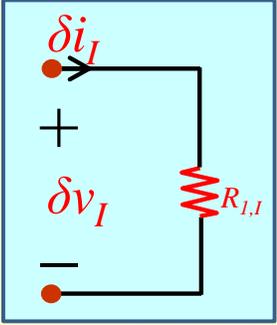

resistor at the DC operating point $Q_I$ is given by,

$$Y_I(s;Q_I) = \frac{\hat{i}_I(s)}{\hat{v}_I(s)} = a_{12}(Q_I) = \frac{1}{\frac{1}{a_{12}(Q_I)}} = \frac{1}{R_{1I}} \quad (8h)$$

where

$$R_{1,I} = 1/a_{12}(Q_I) \quad (8i)$$

From (8h), it is followed that the small-signal admittance function of the *mixed ion-channel* nonlinear resistor is equivalent to a linear resistor. The corresponding small-signal equivalent circuit and a plot of the coefficient $a_{12}(Q_I)$ and resistance $R_{1,I}$ as a function of the DC equilibrium voltage $V_I$ are shown in Fig. 6(a) and Fig. 6(b), respectively. The explicit formulas for computing coefficient $a_{12}(Q_I)$ are given in Table 3 for readers convenience.

### 4.2 Small-signal circuit model of the voltage-sensitive potassium ion-channel memristor

The small-signal circuit model of the *voltage sensitive potassium ion-channel* memristor at an equilibrium point $Q_{K,V}$ [6] on the DC $V_{K,V}$-$I_{K,V}$ curve is derived by defining

$$n = n_{Q_{K,V}} + \delta n \quad (9a)$$

$$v_{K,V} = V_{K,V}(Q_{K,V}) + \delta v_{K,V} \quad (9b)$$

$$i_{K,V} = I_{K,V}(Q_{K,V}) + \delta i_{K,V} \quad (9c)$$

---
[6] The equilibrium point $Q_{K,V}$ at $v_{K,V} = V_{K,V}$ is obtained from (4d) by solving $f(n;V_{K,V}) = 0$ for $n = n_{K,V}$. The explicit formula for $n(V_{K,V})$ is given in Table 4.

Expanding $i_{K,V} = G_{K,V}(n)v_{K,V}$ from (4b) in a Taylor series about the equilibrium point $(N(Q_{K,V}), V_{K,V}(Q_{K,V}))$, we obtain,

$$i_{K,V} = a_{00}(Q_{K,V}) + a_{11}(Q_{K,V})\delta n + a_{12}(Q_{K,V})\delta v_{K,V} + h.o.t.$$
$$= I_{K,V}(Q_{K,V}) + \delta i_{K,V} \quad (9d)$$

where

$$\delta n = n - n_{Q_{K,V}}, \quad \delta v_{K,V} = v_{K,V} - V_{K,V}(Q_{K,V}),$$
$$\delta i_{K,V} = i_{K,V} - I_{K,V}(Q_{K,V}) \quad (9e)$$

and

$$a_{00}(Q_{K,V}) = G_{K,V}(Q_{K,V})V_{K,V}(Q_{K,V}) = I_{K,V}(Q_{K,V}) \quad (9f)$$

$$a_{11}(Q_{K,V}) = V_{K,V}(Q_{K,V})G'_{K,V}(n_{Q_{K,V}}) \quad (9g)$$

$$a_{12}(Q_{K,V}) = G_{K,V}(n_{Q_{K,V}}) \quad (9h)$$

and h.o.t denotes the higher-order terms. Let us linearize the nonlinear equation by neglecting the h.o.t. in (9d), then:

$$\boxed{\delta i_K = a_{11}(Q_{K,V})\delta n + a_{12}(Q_{K,V})\delta v_{K,V}} \quad (9i)$$

Similarly, expanding the state equation $f(n_{K,V}, V_{K,V})$ in (4d) using a Taylor series about the equilibrium point $(n(Q_{K,V}), V_{K,V}(Q_{K,V}))$, we obtain

$$f(n_{Q_{K,V}} + \delta n, V_{K,V}(Q_{K,V}) + \delta v_{K,V})$$
$$= f(n_{Q_{K,V}}, V_{K,V}(Q_{K,V})) + b_{11}(Q_{K,V})\delta n + b_{12}(Q_{K,V})\delta v_{K,V} + h.o.t. \quad (9j)$$

where

$$b_{11}(Q_{K,V}) = \left.\frac{\partial f_n(n, v_{K,V})}{\partial n}\right|_{Q_{K,V}} \quad (9k)$$

$$b_{12}(Q_{K,V}) = \left.\frac{\partial f_N(n, v_{K,V})}{\partial v_{K,V}}\right|_{Q_{K,V}} \quad (9l)$$

Linearizing the nonlinear state equation (9j) by neglecting the h.o.t., we get

$$\boxed{\frac{d(\partial n)}{dt} = b_{11}(Q_{K,V})\delta n + b_{12}(Q_{K,V})\delta v_{K,V}} \quad (9m)$$

Taking Laplace transform of (9i) and (9m), we obtain

$$\hat{i}_{K,V}(s) = a_{11}(Q_{K,V})\hat{n}(s) + a_{12}(Q_{K,V})\hat{v}_{K,V}(s) \quad (9n)$$

$$s\,\hat{n}(s) = b_{11}(Q_{K,V})\hat{n}(s) + b_{12}(Q_{K,V})\hat{v}_{K,V}(s) \quad (9o)$$

Solving (9o) for $\hat{n}(s)$ and substituting the result into (9n), we obtain the following admittance $Y_{K,V}(s; Q_{K,V})$ of the small-signal equivalent circuit of the *voltage sensitive potassium ion-channel* memristor at equilibrium point $Q_{K,V}$:

$$Y_{K,V}(s; Q_{K,V}) = \frac{\hat{i}_{K,V}(s)}{\hat{v}_{K,V}(s)}$$

$$= \left[\frac{1}{\dfrac{s}{a_{11}(Q_{K,V})b_{12}(Q_{K,V})} - \dfrac{b_{11}(Q_{K,V})}{a_{11}(Q_{K,V})b_{12}(Q_{K,V})}} + \frac{1}{a_{12}(Q_{K,V})}\right] \quad (9p)$$

$$\boxed{Y_{K,V}(s; Q_{K,V}) = \left(\frac{1}{(sL_{K,V} + R_{1K,V})} + \frac{1}{R_{2K,V}}\right)} \quad (9q)$$

where

$$L_{K,V} \triangleq \frac{1}{a_{11}(Q_{K,V})b_{12}(Q_{K,V})} \quad (9r)$$

$$R_{1K,V} \triangleq -\frac{b_{11}(Q_{K,V})}{a_{11}(Q_{K,V})b_{12}(Q_{K,V})} \quad (9s)$$

$$R_{2K,V} \triangleq \frac{1}{a_{12}(Q_{K,V})} \quad (9t)$$

It follows from (9r)-(9t) that the small-signal admittance function of the first-order voltage sensitive *potassium ion-channel memristor* is equivalent to the serial connection of an inductor and a resistor in parallel with another resistor as shown in Fig. 7. The corresponding coefficients $a_{11}$, $a_{12}$, $b_{11}$, $b_{12}$ and *inductance* $L_{K,V}$, *resistance* $R_{1K,V}$ and *resistance* $R_{2K,V}$ as a function of the DC equilibrium voltage $V_{K,V}$ are shown in Fig. 8 and Fig. 9, respectively. For the readers convenience, the explicit formulas for computing the coefficients $a_{11}(Q_{K,V})$, $a_{12}(Q_{K,V})$, $b_{11}(Q_{K,V})$, $b_{12}(Q_{K,V})$ and $L_{K,V}$, $R_{1K,V}$, $R_{2K,V}$ are summarized in Table 4.

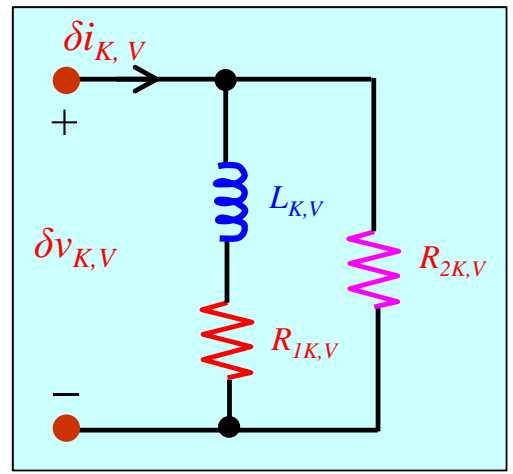

**Fig. 7.** Small-signal equivalent circuit model of the *voltage-sensitive potassium ion-channel* memristor about the DC equilibrium point $Q_{K,V}(V_{K,V}, I_{K,V})$.

**Table 4:** Explicit formulas for computing the coefficients $a_{11}(Q_{K,V})$, $a_{12}(Q_{K,V})$, $b_{11}(Q_{K,V})$, $b_{12}(Q_{K,V})$ and $L_{K,V}$, $R_{1K,V}$, $R_{2K,V}$ of the *voltage sensitive potassium ion-channel* memristor.

$$a_{11}(Q_{K,V}) = 4g_{K,V} n(V_{K,V})^3 V_{K,V}$$

$$a_{12}(Q_{K,V}) = g_{K,V} n(V_{K,V})^4$$

$$n(V_{K,V}) = \frac{\alpha_n(V_{K,V})}{\alpha_n(V_{K,V}) + \beta_n(V_{K,V})}$$

$$\alpha_n(V_{K,V}) = \frac{0.01(V_{K,V} + E_K + 20)}{1 - e^{-0.1(V_{K,V} + E_K + 20)}}$$

$$L_{K,V} = \frac{1}{a_{11}(Q_{K,V}) b_{12}(Q_{K,V})}$$

$$R_{1K,V} = -\frac{b_{11}(Q_{K,V})}{a_{11}(Q_{K,V}) b_{12}(Q_{K,V})}$$

$$\beta_n(V_{K,V}) = 0.125 e^{\left(\frac{-(V_{K,V} + E_K + 30)}{80}\right)}$$

$$b_{11}(Q_{K,V}) = -\lambda_n \left[\alpha_n(V_{K,V}) + \beta_n(V_{K,V})\right]$$

$$R_{2K,V} = \frac{1}{a_{12}(Q_{K,V})}$$

$$b_{12}(Q_{K,V}) = \lambda_n \left[\frac{0.01 - 0.1\alpha_n(V_{K,V}) e^{-0.1(V_{K,V} + E_K + 20)}}{(1 - e^{-0.1(V_{K,V} + E_K + 20)})}\right] (1 - n(V_{K,V})) + \frac{\beta_n(V_{K,V})}{80} n(V_{K,V})$$

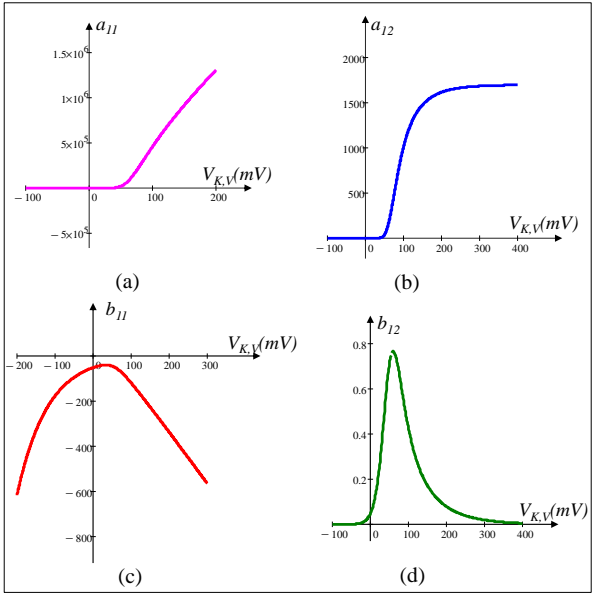

**Fig. 8.** Plot of coefficients (a) $a_{11}$ (b) $a_{12}$ (c) $b_{11}$ and (d) $b_{12}$ of the voltage-sensitive potassium ion-channel memristor as a function of the DC equilibrium voltage $V_{K,V}$.

### 4.3 Small-signal circuit model of the calcium-sensitive potassium ion-channel memristor

The small-signal circuit model of the *calcium-sensitive potassium-channel* memristor at an equilibrium point $Q_{K,Ca}$ [7] in the DC $V_{K,Ca}$-$I_{K,Ca}$ curve is derived by defining

$$Ca = Ca_{Q_{K,Ca}} + \delta Ca \tag{10a}$$

---
[7] The equilibrium point $Q_{K,Ca}$ at $v_{K,Ca} = V_{K,Ca}$ is obtained from (5a) by solving $f(Ca; V_{K,Ca}) = 0$ for $Ca = Ca_{K,Ca}$. The explicit formula for $Ca(V_{K,Ca})$ is given in Table 5.

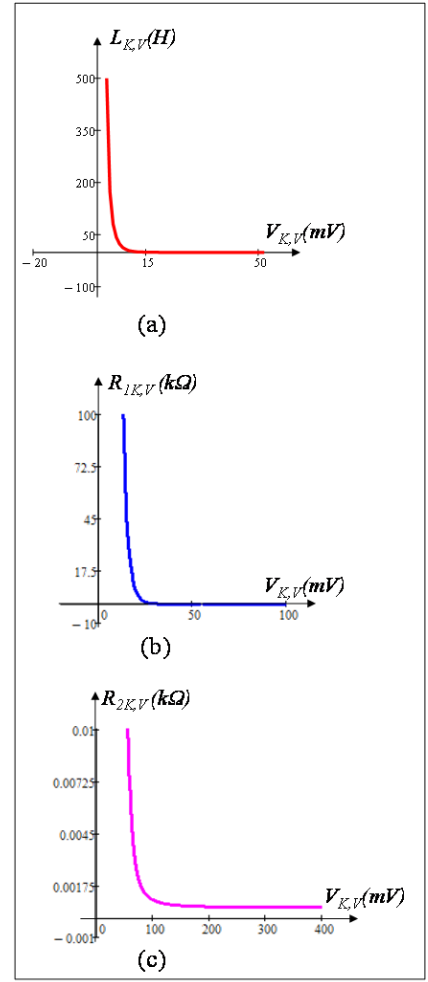

**Fig. 9.** (a) Inductance $L_{K,V}$ (b) resistance $R_{1K,V}$ and (c) resistance $R_{2K,V}$ of the *voltage-sensitive potassium ion-channel* memristor as a function of DC equilibrium voltage $V_{K,V}$.

$$v_{K,Ca} = V_{K,Ca}(Q_{Ca}) + \delta v_{K,Ca} \quad (10b)$$

$$i_{K,Ca} = I_{K,Ca}(Q_{K,Ca}) + \delta i_{K,Ca} \quad (10c)$$

Expanding $i_{K,Ca} = G_{K,Ca}(Ca) v_{K,Ca}$ from (5a) in a Taylor series about the equilibrium point $(Ca(Q_{K,Ca}), V_{Ca}(Q_{K,Ca}))$, we obtain

$$i_{K,Ca} = a_{00}(Q_{K,Ca}) + a_{11}(Q_{K,Ca}) \delta Ca + a_{12}(Q_{K,Ca}) \delta v_{K,Ca} + h.o.t.$$
$$= I_{K,Ca}(Q_{Ca}) + \delta i_{K,Ca}$$

(10d)

where

$$\delta Ca = Ca - Ca_{Q_{K,Ca}}, \quad \delta v_{K,Ca} = v_{K,Ca} - V_{K,Ca}(Q_{K,Ca}),$$
$$\delta i_{K,Ca} = i_{K,Ca} - I_{K,Ca}(Q_{K,Ca}), \quad (10e)$$

and

$$a_{00}(Q_{K,Ca}) = G_{Ca}(Q_{K,Ca}) V_{Ca}(Q_{K,Ca}) = I_{K,Ca}(Q_{K,Ca}) \quad (10f)$$

$$a_{11}(Q_{K,Ca}) = V_{K,Ca}(Q_{K,Ca}) G'_{K,Ca}\left(Ca_{Q_{K,Ca}}\right) \quad (10g)$$

$$a_{12}(Q_{K,Ca}) = G_{K,Ca}\left(Ca_{Q_{K,Ca}}\right) \quad (10h)$$

and h.o.t denotes the higher-order terms. Let us linearize the nonlinear equation by neglecting the h.o.t. in (10d) then:

$$\boxed{\delta i_{K,Ca} = a_{11}(Q_{K,Ca}) \delta Ca + a_{12}(Q_{K,Ca}) \delta v_{K,Ca}} \quad (10i)$$

Similarly, expanding the state equation $f(Ca_{K,Ca}, V_{K,Ca})$ of (5d) in a Taylor series about the equilibrium point $(Ca(Q_{,Q,K,Ca}), V_{Ca}(Q_{,K,ca}))$, we obtain

$$f(Ca_{Q_{K,Ca}} + \delta Ca, V_{K,Ca}(Q_{K,Ca}) + \delta v_{K,Ca})$$
$$= f(Ca_{Q_{K,Ca}}, V_{Ca}(Q_{K,Ca})) + b_{11}(Q_{K,Ca}) \delta Ca + b_{12}(Q_{K,Ca}) \delta v_{K,Ca} + h.o.t.$$

(10j)

where

$$b_{11}(Q_{K,Ca}) = \left.\frac{\partial f(Ca, v_{K,Ca})}{\partial Ca}\right|_{Q_{K,Ca}} \quad (10k)$$

$$b_{12}(Q_{K,Ca}) = \left.\frac{\partial f(Ca, v_{K,Ca})}{\partial v_{K,Ca}}\right|_{Q_{K,Ca}} \quad (10l)$$

Linearizing the nonlinear state equation (10j) by neglecting the h.o.t., we get

$$\boxed{\frac{d(\partial Ca)}{dt} = b_{11}(Q_{K,Ca}) \delta Ca + b_{12}(Q_{K,Ca}) \delta v_{K,Ca}} \quad (10m)$$

Taking Laplace transform of (10i) and (10m), we obtain

$$\hat{i}_{K,Ca}(s) = a_{11}(Q_{K,Ca}) C\hat{a}(s) + a_{12}(Q_{K,Ca}) \hat{v}_{Ca}(s) \quad (10n)$$

$$s \, C\hat{a}(s) = b_{11}(Q_{K,Ca}) C\hat{a}(s) + b_{12}(Q_{K,Ca}) \hat{v}_{K,Ca}(s) \quad (10o)$$

Solving (10o) for $C\hat{a}(s)$ and substituting the result into (10n), we obtain the following admittance $Y_{K,Ca}(s; Q_{K,Ca})$ of the small-signal equivalent circuit of the *calcium sensitive potassium ion-channel* memristor at equilibrium point $Q_{K,Ca}$:

$$Y_{K,Ca}(s; Q_{K,Ca}) = \frac{\hat{i}_{K,Ca}(s)}{\hat{v}_{K,Ca}(s)}$$

$$= \left[ \frac{1}{\dfrac{s}{a_{11}(Q_{K,Ca}) b_{12}(Q_{K,Ca})} - \dfrac{b_{11}(Q_{K,Ca})}{a_{11}(Q_{K,Ca}) b_{12}(Q_{K,Ca})}} + \frac{1}{\dfrac{1}{a_{12}(Q_{K,Ca})}} \right]$$

(10p)

$$\boxed{Y_{Ca}(s; Q_{Ca}) = \left( \frac{1}{\left(sL_{K,Ca} + R_{1K,Ca}\right)} + \frac{1}{R_{2K,Ca}} \right)} \quad (10q)$$

where

$$L_{K,Ca} \triangleq \frac{1}{a_{11}(Q_{K,Ca}) b_{12}(Q_{K,Ca})} \quad (10r)$$

$$R_{1K,Ca} \triangleq -\frac{b_{11}(Q_{K,Ca})}{a_{11}(Q_{K,Ca}) b_{12}(Q_{K,Ca})} \quad (10s)$$

$$R_{2K,Ca} \triangleq \frac{1}{a_{12}(Q_{K,Ca})} \quad (10t)$$

It follows from (10r)-(10t) that the small-signal admittance function of the first-order *calcium-sensitive potassium ion-channel memristor* is equivalent to the serial connection of an inductor and a resistor in parallel with another resistor as shown in Fig. 10. The corresponding coefficients $a_{11}$, $a_{12}$, $b_{11}$, $b_{12}$ and *inductance* $L_{K,Ca}$, *resistance* $R_{1K,Ca}$, and *resistance* $R_{2K,Ca}$ as a function of the DC equilibrium voltage $V_{K,Ca}$ are shown in Fig. 11 and Fig. 12, respectively. For the readers convenience, the explicit formulas for computing the coefficients $a_{11}(Q_{,K,Ca})$, $a_{12}(Q_{K,Ca})$, $b_{11}(Q_{K,Ca})$, $b_{12}(Q_{K,Ca})$ and $L_{K,Ca}$, $R_{1K,Ca}$, $R_{2K,Ca}$ are summarized in Table 5.

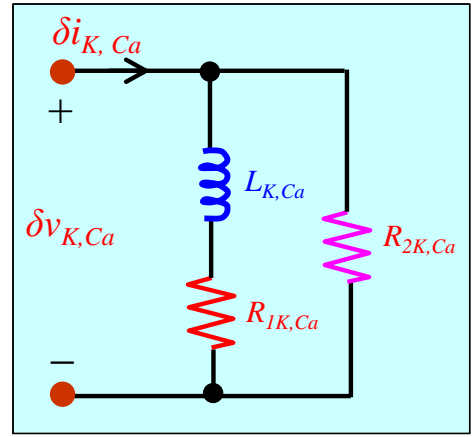

**Fig. 10.** Small-signal equivalent circuit model of the *calcium-sensitive potassium ion-channel memristor* about the DC equilibrium point $Q_{K,Ca}$ $(V_{K,Ca}, I_{K,Ca})$.

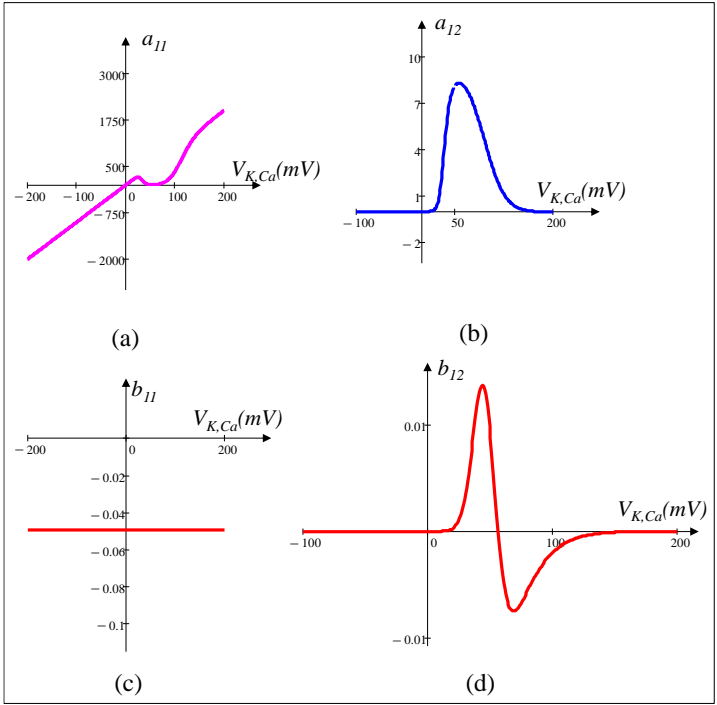

**Fig. 11.** Plot of coefficients (a) $a_{11}$ (b) $a_{12}$ (c) $b_{11}$ and (d) $b_{12}$ of the *calcium-sensitive potassium ion-channel* memristor as a function of the DC equilibrium voltage $V_{K,Ca}$.

### 4.4 Small-signal circuit model of the memrisive Chay model

Let us replace the small signal models of the *voltage-sensitive ion-channel nonlinear resistor*, *voltage sensitive-potassium ion-channel memristor* and *calcium-sensitive ion-channel memristor* at DC operating voltage $V_I = V - E_I$, $V_{K,V} = V - E_K$ and $V_{K,Ca} = V - E_K$ respectively to the memristive Chay model in Fig. 1(b). The equivalent small signal circuit composed of capacitor, inductors and resistors are shown in Fig. 13. The admittance $Y(s; Vm(Q))$ at the equilibrium point $Q$ at $V = Vm(Q)$ is given by

$$Y(s; V_m(Q)) = sC_m + \frac{1}{sL_{K,V} + R_{1K,V}} + \frac{1}{sL_{K,Ca} + R_{1K,ca}} + \frac{1}{R_{1,I}} + \frac{1}{R_{2K,V}} + \frac{1}{R_{2K,Ca}} + G_L \quad (11)$$

The circuit element $R_{1,I}$ is obtained by calculating the small signal model of the voltage-sensitive mixed ion-channel nonlinear resistor from Table 3 at equilibrium voltage $V_m(Q)$ where $V_m(Q) = V_I + E_I$. Similarly, $L_{K,V}$, $R_{1K,V}$, and $R_{2K,V}$ are calculated from the small-signal equivalent circuit of the *voltage sensitive potassium ion-channel memristor* from Table 4, and $L_{K,Ca}$, $R_{1K,Ca}$, and $R_{2K,Ca}$ are calculated from the small signal equivalent circuit of the *calcium-sensitive potassium ion channel memristor* from Table 5 at equilibrium voltage $V_m(Q)$ respectively. Note that $V_{K,V} + E_K$ and $V_{K,Ca} + E_K$ must be replaced by $V_m(Q)$ in Table 4 and Table 5 by the small signal model of the *voltage-sensitive potassium ion channel memristor* and *calcium-sensitive potassium ion channel memristor*, respectively.

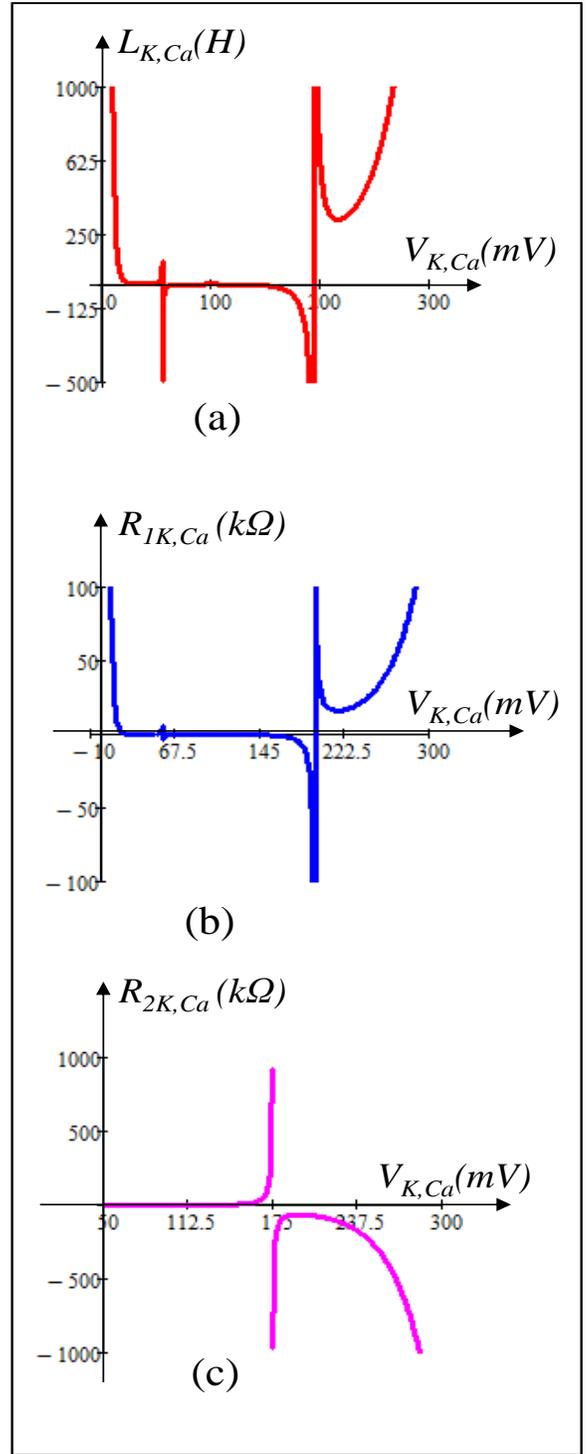

**Fig. 12.** (a) Inductance $L_{K,Ca}$ (b) resistance $R_{1K,Ca}$ and (c) resistance $R_{2K,Ca}$ of the *calcium-sensitive potassium ion-channel memristor* as a function of DC equilibrium voltage $V_{K,Ca}$.

**Table 5:** Explicit formulas for computing the coefficients $a_{11}(Q_{K,Ca})$, $a_{12}(Q_{K,Ca})$, $b_{11}(Q_{K,Ca})$, $b_{12}(Q_{K,Ca})$ and $L_{K,Ca}$, $R_{1K,Ca}$, $R_{2K,Ca}$ of the *calcium-sensitive potassium ion-channel* memristor.

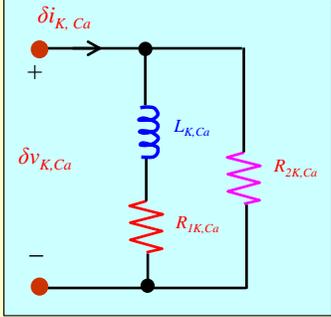

$$m(V_{K,Ca}) = \frac{0.1(V_{K,Ca} + E_K + 25)}{0.1(V_{K,Ca} + E_K + 25) + 4\left(1 - e^{-0.1(V_{K,Ca}+E_K+25)}\right)e^{\left(\frac{-(V_{K,Ca}+E_K+50)}{18}\right)}}$$

$$h(V_{K,Ca}) = \frac{\left(1 + e^{-0.1(V_{K,Ca}+E_K+20)}\right)0.07 e^{\left(\frac{-(V_{K,Ca}+E_K+50)}{20}\right)}}{\left(1 + e^{-0.1(V_{K,Ca}+E_K+20)}\right)0.07 e^{\left(\frac{-(V_{K,Ca}+E_K+50)}{20}\right)} + 1}$$

$$Ca(V_{K,Ca}) = \frac{-m(V_{K,Ca})^3 h(V_{K,Ca})(V_{K,Ca} + E_K - E_{Ca})}{k_{Ca}}$$

$$a_{11}(Q_{K,Ca}) = \frac{g_{K,Ca} V_{K,Ca}}{(1 + Ca(V_{K,Ca}))^2}$$

$$a_{12}(Q_{K,Ca}) = g_{K,Ca} \frac{Ca(V_{K,Ca})}{1 + Ca(V_{K,Ca})}$$

$$b_{11}(Q_{K,Ca}) = -\rho k_{Ca}$$

$$b_{12}(Q_{K,Ca}) = -\rho\left[m(V_{K,Ca})^3 h(V_{K,Ca}) + m(V_{K,Ca})^3(V_{K,Ca} + E_K - E_{Ca})\frac{dh(V_{K,Ca})}{dV_{K,Ca}} + h(V_{K,Ca})(V_{K,Ca} + E_K - E_{Ca})\frac{dm(V_{K,Ca})^3}{dV_{K,Ca1}}\right]$$

$$h_d(V_{K,Ca}) = \frac{dh(V_{K,Ca})}{dV_{K,Ca}} = -h(V_{K,Ca})^2 \frac{\frac{0.07}{20}\left(1 + e^{-0.1(V_{K,Ca}+E_K+20)}\right)e^{\left(\frac{-(V_{K,Ca}+E_K+50)}{20}\right)} + 0.007 e^{\left(\frac{-(V_{K,Ca}+E_K+50)}{20}\right)}e^{-0.1(V_{K,Ca}+E_K+20)}}{\left(\left(1 + e^{-0.1(V_{K,Ca}+E_K+20)}\right)0.07 e^{\left(\frac{-(V_{K,Ca}+E_K+50)}{20}\right)}\right)^2}$$

$$m_d(V_{K,Ca}) = \frac{dm^2(V_{K,Ca})}{dV_{K,Ca}} = 3m^2(V_{K,Ca})\frac{dm(V_{K,Ca})}{dV_I} = 3m(V_{K,Ca})^2 m_{dd}(V_{K,Ca})$$

$$m_{dd}(V_{K,Ca}) = \frac{-40 m(V_{K,Ca})^2}{(V_{K,Ca} + E_K + 25)^2}\left[(V_{K,Ca} + E_K + 25) m_{ddd}(V_{K,Ca}) - \left(1 - e^{-0.1(V_{K,Ca}+E_K+25)}\right)e^{\left(\frac{-(V_{K,Ca}+E_K+50)}{18}\right)}\right]$$

$$m_{ddd}(V_{K,Ca}) = \frac{-1}{18}\left(1 - e^{-0.1(V_{K,Ca}+E_K+25)}\right)e^{\left(\frac{-(V_{K,Ca}+E_K+50)}{18}\right)} + 0.1 e^{\left(\frac{-(V_{K,Ca}+E_K+50)}{18}\right)}e^{-0.1(V_{K,Ca}+E_K+25)}$$

$$L_{K,Ca} = \frac{1}{a_{11}(Q_{K,Ca}) b_{12}(Q_{K,Ca})}$$

$$R_{1K,Ca} = -\frac{b_{11}(Q_{K,Ca})}{a_{11}(Q_{K,Ca}) b_{12}(Q_{K,Ca})}$$

$$R_{2K,Ca} = \frac{1}{a_{12}(Q_{K,Ca})}$$

*4.4.1. Frequency response*

A convenient way to find the total admittance $Y(s; V_m(Q))$ by recasting (11) into a rational function of the complex frequency $s$, is as follows:

$$Y(s; V_m(Q)) = \frac{b_3 s^3 + b_2 s^2 + b_1 s + b_0}{a_2 s^2 + a_1 s + a_0} \quad (12a)$$

where the explicit formula for computing the coefficients $b_3$, $b_2$, $b_1$, $b_0$ $a_2$, $a_1$, and $a_0$ are summarized in Table 6.

Substituting $s = i\omega$ in (12a), we obtain the following small-signal admittance functions at the equilibrium voltage $V_m(Q)$:

$$Y(i\omega; V_m(Q)) = \frac{(b_0 - b_2\omega^2)(a_0 - a_2\omega^2) + a_1\omega^2(b_1 - b_3\omega^2)}{(a_0 - a_2\omega^2)^2 + (a_1\omega)^2}$$

$$+ i\omega\left[\frac{(b_1 - b_3\omega^2)(a_0 - a_2\omega^2) - a_1(b_0 - b_2\omega^2)}{(a_0 - a_2\omega^2)^2 + (a_1\omega)^2}\right]$$

(12b)

The corresponding *real part* $Re\ Y(i\omega; V_m(Q))$ and *imaginary part* $Im\ Y(i\omega; V_m(Q))$ from (12b) is given by,

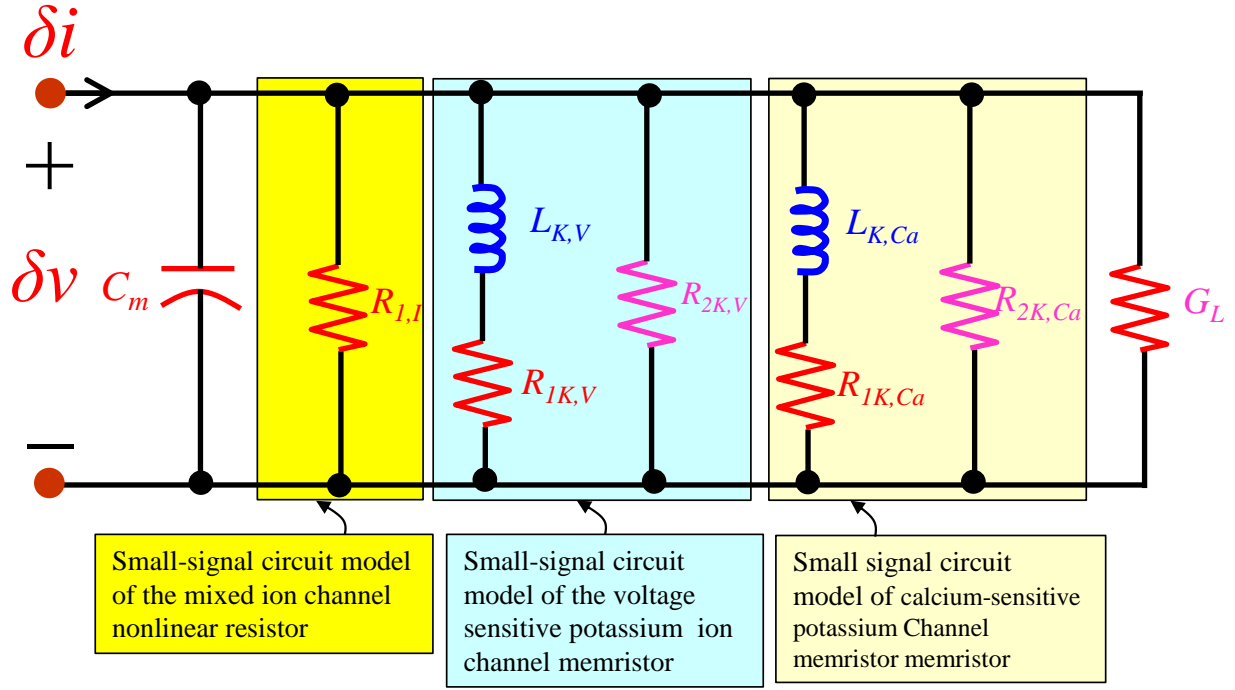

**Fig. 13.** Small-signal equivalent circuit model of the memristive Chay model.

Table 6: Explicit formulas for computing the coefficients of $Y(s;V_m(Q))$.

$$b_3 = L_{K,V} L_{K,Ca} R_{1,I} R_{2K,V} R_{2K,Ca} C_m$$

$$b_2 = \left( L_{K,V} R_{1K,ca} + L_{K,Ca} R_{1K,V} \right) R_{1,I} R_{2K,V} R_{2K,Ca} C_m + L_{K,V} L_{K,Ca} R_{2K,V} R_{2K,Ca}$$
$$+ L_{K,V} L_{K,Ca} R_{1,I} R_{2K,Ca} + L_{K,V} L_{K,Ca} R_{1,I} R_{2K,V} + L_{K,V} L_{K,Ca} R_{1,I} R_{2K,V} R_{2K,Ca} G_L$$

$$b_1 = R_{1,I} R_{1K,V} R_{1K,ca} R_{2K,V} R_{2K,Ca} C_m + \left( L_{K,Ca} R_{1,I} R_{2K,V} R_{2K,Ca} \right) + \left( L_{K,V} R_{1,I} R_{2K,V} R_{2K,Ca} \right)$$
$$+ \left( L_{K,V} R_{1K,ca} + L_{K,Ca} R_{1K,V} \right) R_{2K,V} R_{2K,Ca} + \left( L_{K,V} R_{1K,ca} + L_{K,Ca} R_{1K,V} \right) R_{1,I} R_{2K,Ca}$$
$$+ \left( L_{K,V} R_{1K,ca} + L_{K,Ca} R_{1K,V} \right) R_{1,I} R_{2K,V} + \left( L_{K,V} R_{1K,ca} + L_{K,Ca} R_{1K,V} \right) R_{1,I} R_{2K,V} R_{2K,Ca} G_L$$

$$b_0 = R_{1,I} R_{1K,ca} R_{2K,V} R_{2K,Ca} + R_{1,I} R_{1K,V} R_{2K,V} R_{2K,Ca} + R_{1K,V} R_{1K,ca} R_{2K,V} R_{2K,Ca}$$
$$+ R_{1,I} R_{1K,V} R_{1K,ca} R_{2K,Ca} + R_{1,I} R_{1K,V} R_{1K,ca} R_{2K,V} + R_{1,I} R_{1K,V} R_{1K,ca} R_{2K,V} R_{2K,Ca} G_L$$

$$a_2 = L_{K,V} L_{K,Ca} R_{1,I} R_{2K,V} R_{2K,Ca}$$

$$a_1 = \left( L_{K,V} R_{1K,ca} + L_{K,Ca} R_{1K,V} \right) R_{1,I} R_{2K,V} R_{2K,Ca}$$

$$a_0 = R_{1,I} R_{1K,V} R_{1K,ca} R_{2K,V} R_{2K,Ca}$$

$$Y(s; V_m(Q)) = \frac{b_3 s^3 + b_2 s^2 + b_1 s + b_0}{a_2 s^2 + a_1 s + a_0}$$

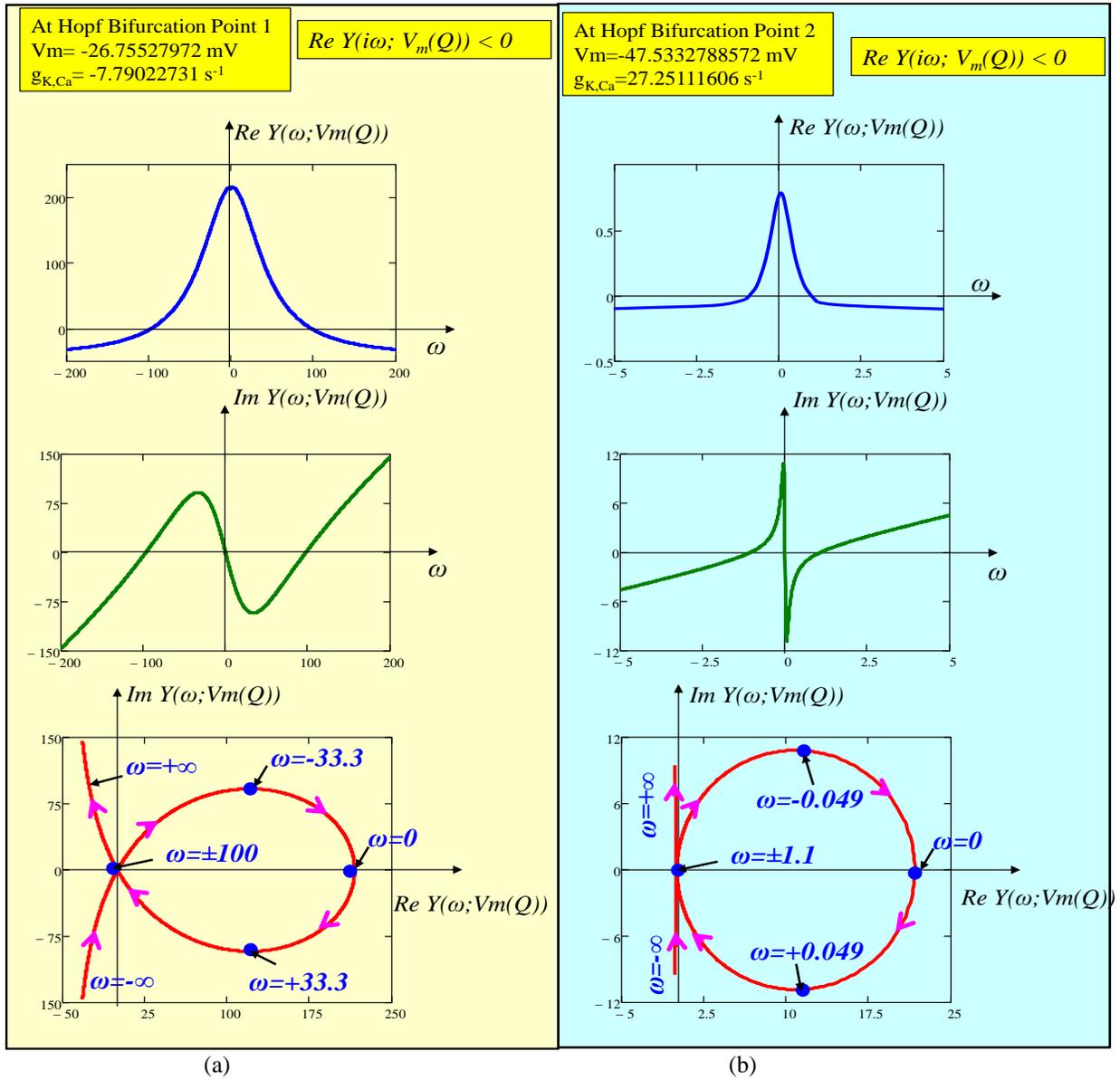

**Fig. 14.** Small-signal admittance frequency response and Nyquist plot of the memristive Chay model at (a) $V_m= -26.75527972$ mV(resp., $g_{K,Ca}= -7.79022731$ $s^{-1}$) and (b) $V_m=-47.5332788572$ mV(resp., $g_{K,Ca}=27.25111606$ $s^{-1}$). Observe that $ReY(i\omega;V_m(Q))<0$ at the two Hopf-bifurcation points.

$$\operatorname{Re} Y(i\omega;V_m(Q)) = \left[ \frac{(b_0 - b_2\omega^2)(a_0 - a_2\omega^2) + a_1\omega^2(b_1 - b_3\omega^2)}{(a_0 - a_2\omega^2)^2 + (a_1\omega)^2} \right]$$

$$\operatorname{Im} Y(i\omega;V_m(Q)) = \omega \left[ \frac{(b_1 - b_3\omega^2)(a_0 - a_2\omega^2) - a_1(b_0 - b_2\omega^2)}{(a_0 - a_2\omega^2)^2 + (a_1\omega)^2} \right]$$

(12c)

Fig. 14(a) and Fig. 14(b) show $ReY(i\omega; V_m(Q))$ versus $\omega$, $Im Y(i\omega; V_m(Q))$ versus $\omega$, and the Nyquist plot $Im Y(i\omega; V_m(Q))$ versus $Re Y(i\omega; V_m(Q))$ at the DC equilibrium voltage $V_m= -26.75527972$ mV(resp., $g_{K,Ca}= -7.79022731$ $s^{-1}$), and $V_m=-47.5332788572$ mV(resp., $g_{K,Ca}=27.25111606$ $s^{-1}$), respectively. Observe from Fig. 14(a) and Fig. 14(b) that $ReY(i\omega; V_m(Q))<0$, thereby confirming the memristive Chay model is a *locally active* at the above two equilibria. Our extensive numerical computations show that the two DC equilibria coincide with two-Hopf bifurcation points which is mechanism to generate voltage oscillation, spikes, chaos and bursting in excitable cells. We will discuss these two bifurcation points in next section with pole-zeros and Eigen values diagram.

*4.4.2. Pole-zero diagram of the small-signal admittance function $Y(s; V_m(Q))$ and Eigen values of the Jacobean Matrix*

The location of the poles and zeros of the small signal admittance function $Y(s; V_m(Q))$ of (12a) is computed by factorizing it's denominator and numerators as

$$Y(s; V_m(Q)) = \frac{k(s-z_1)(s-z_2)(s-z_3)}{(s-p_1)(s-p_2)} \quad (13)$$

The poles of the small-signal admittance function $Y(s; V_m(Q))$ as a function of the current $V_m$ over $-200\ mV < V_m < 200\ mV$ is shown in Fig. 15. Observe from Fig. 15(a) and Fig. 15(b) that the two poles $Re(p_1)$, $Re(p_2)$ are negative and $Im(p_1)$, $Im(p_2)$ are always zero respectively for the given input $V_m$, which confirm the two poles of the admittance function has no complex frequency.

Fig. 16(a) shows the *Nyquist plot*, i.e. loci of the imaginary part $Im(z_i)$ versus the real part $Re(z_i)$ of the zeros as a function of the input voltage $V_m$ over the interval $-200\ mV \leq V_m \leq 200mV$. Observe that the real part of the two zeros $z_2$ and $z_3$ are zero at $V_m = -26.75527972mV$ (resp. $g_{kCa} = -7.79022731\ s^{-1}$) and $V_m = -47.5332788572\ mV$ (resp. $g_{kCa} = 27.25111606\ s^{-1}$) respectively. The corresponding points when $Re(z_i)=0$ are known as Hopf bifurcation points in bifurcation theory. Fig. 16(b) and Fig. 16(c) show the zoomed version of Fig, 16(a) near to the two bifurcation points respectively. It is also observed that the $Re(z_2)$ and $Re(z_3)$ are always positive in

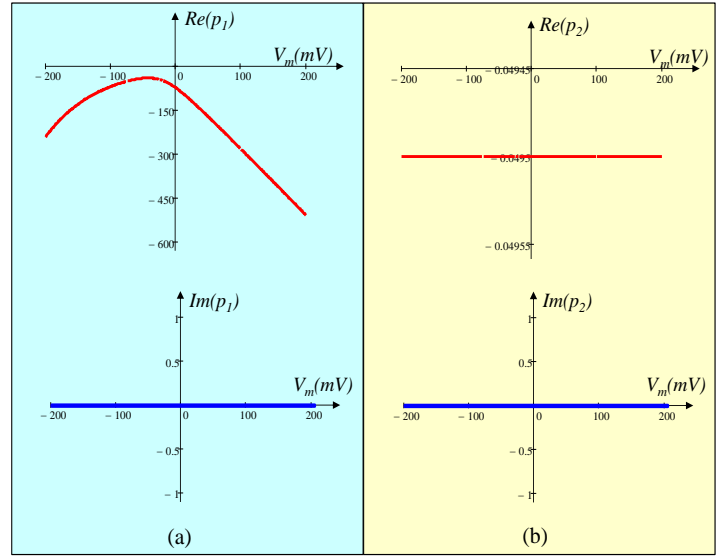

**Fig.15.** Pole diagram of the small-signal admittance function $Y(s; V_m(Q))$ as a function of $V_m$ over $-200\ mV < V_m < 200\ mV$ (a) Top and bottom figures are the plot of the real part of the pole 1 $Re(p_1)$ and Imaginary part of pole 1 $Im(p_1)$ respectively. (b) Top and bottom figures are the plot of the real part of the pole 2 $Re(p_2)$ and Imaginary part of pole 2 $Im(p_2)$ respectively.

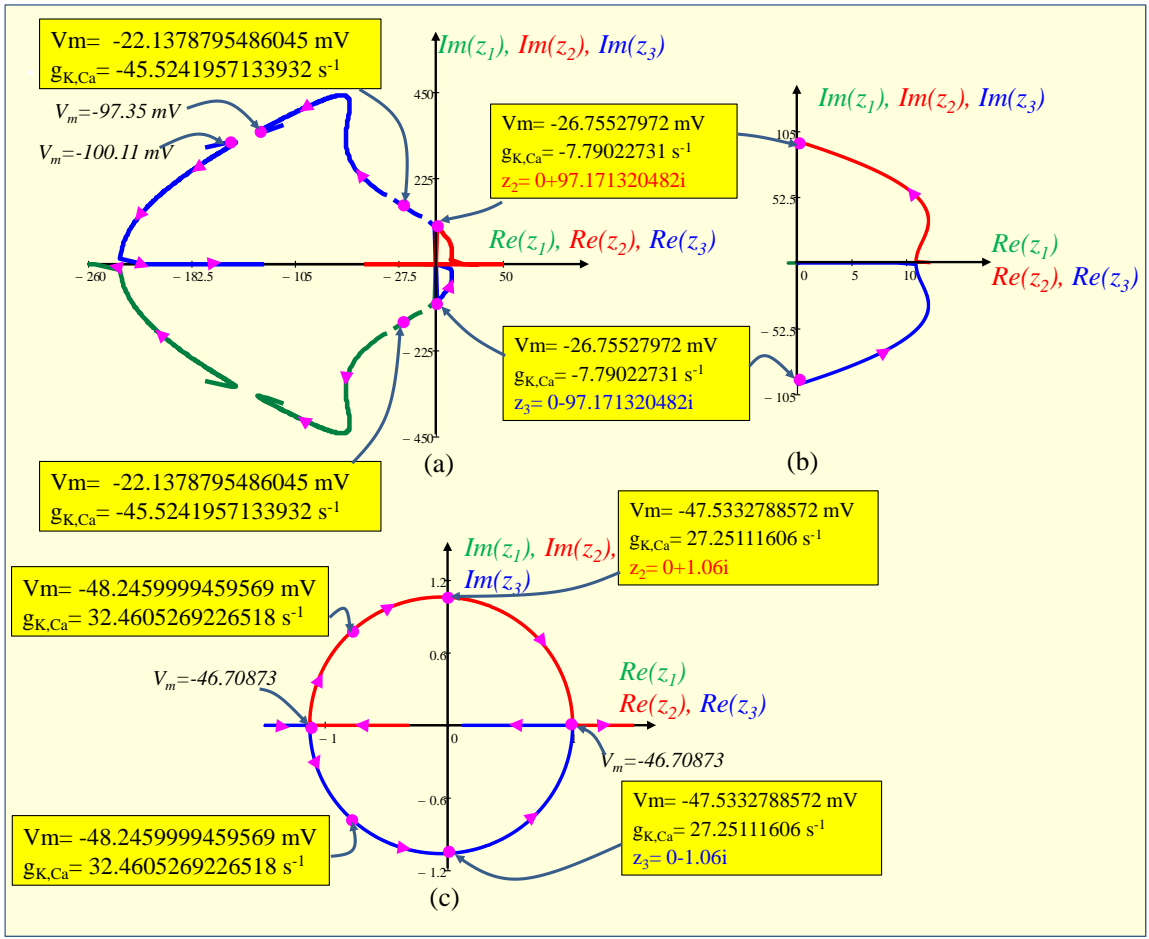

**Fig. 16.** Zeros diagram of the small-signal admittance function $Y(s; V_m(Q))$ (a) Nyquist plot of the zeros $z_1$, $z_2$, $z_3$ in $Im(z_i)$ versus $Re(z_i)$ plane (b) Nyquist plot near the Hopf-bifurcation point 1, $V_m = -26.75527972$ (resp. $g_{kCa} = -7.79022731\ s^{-1}$). (c) Nyquist plot near the Hopf-bifurcation point 2, $V_m = -47.5332788572\ mV$ (resp. $g_{kCa} = 27.25111606\ s^{-1}$).

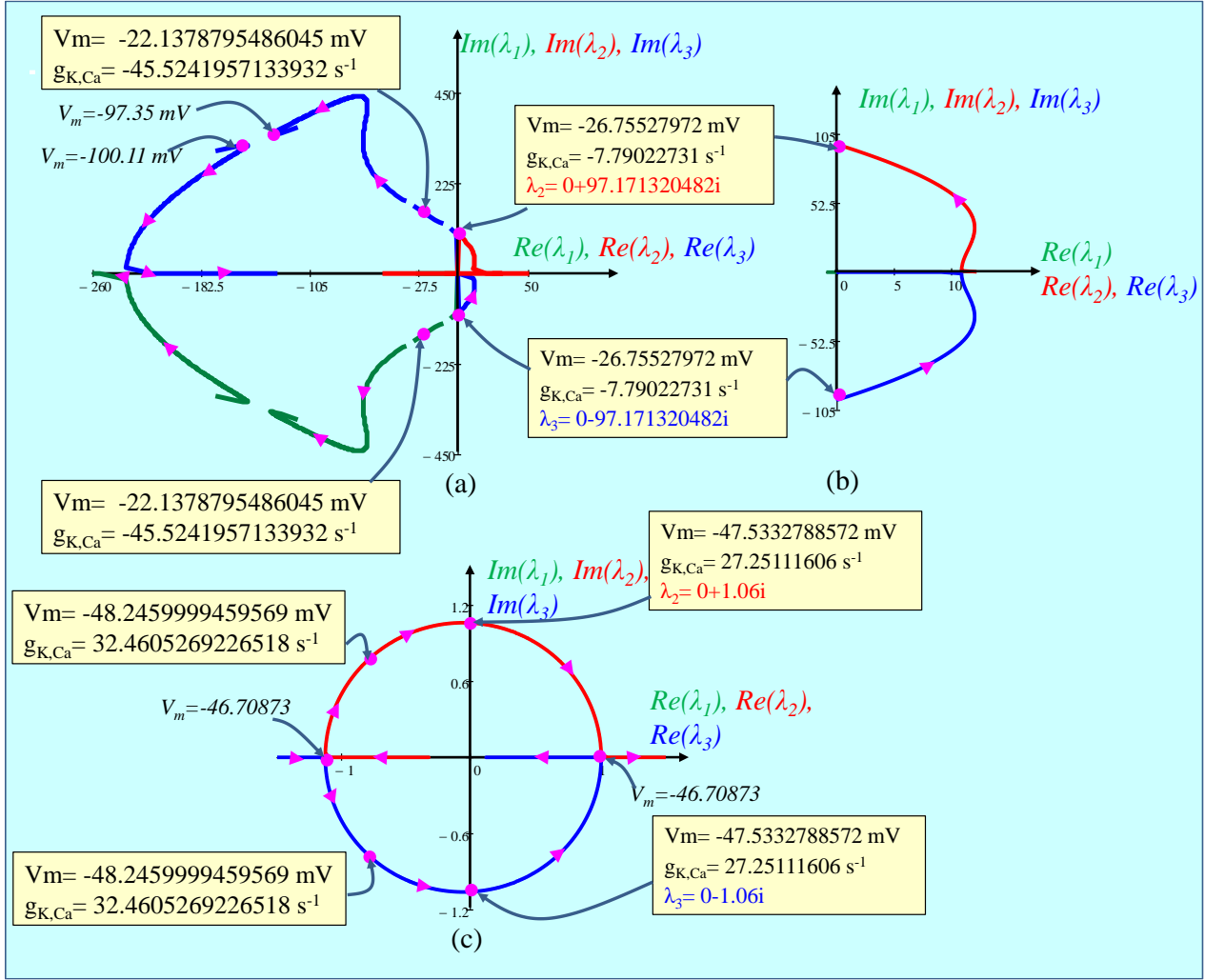

**Fig. 17.** Plot of the loci of the Eigen values of the Jacobean Matrix (a) Nyquist plot of the Eigen values $\lambda_1, \lambda_2, \lambda_3$ in $Im(\lambda_i)$ versus $Re(\lambda_i)$ plane (b) Nyquist plot near the Hopf-bifurcation point 1, $V_m= -26.75527972$ (resp. $g_{kCa}=-7.79022731\ s^{-1}$). (c) Nyquist plot near the Hopf-bifurcation point 2, $V_m= -47.5332788572\ mV$ (resp. $g_{kCa} =27.25111606\ s^{-1}$). Our numerical computations confirm the zeros of the admittance functions $Y(s; V_m(Q))$ obtained in Fig. 16 is identical to the Eigen values of the Jacobean matrix.

open-half plane between the bifurcation points $-26.75527972 > V_m > -47.5332788572\ mV$ (resp. $-7.79022731\ s^{-1} < g_{KCa} < 27.25111606\ s^{-1}$). Observe from Fig. 17 that the Eigen values computed from the Jacobean matrix are identical to the results obtained in Fig. 16 from the zeros of the admittance functions $Y(s; V_m(Q))$ as expected from the Chua theory [Chua et al., 1987, 2012a, 2012b].

## 5. Local Activity, Edge of Chaos and Hopf-Bifurcation in excitable Cells (Memristive Chay Model)

Local Activity and edge of chaos are powerful mathematical theory to predict whether the nonlinear system exhibits complexity or not [Chua, et al., 1987; Chua, 1998; Sah, et al., 2015]. This section presents an extensive numerical simulation using the principle of local activity, edge of chaos and Hopf-bifurcation theorem to predict the mechanism to generate the complicated electrical signals in memristive Chay model.

### 5.1 Locally active regime

We performed comprehensive numerical simulations within the range of the DC equilibrium voltage $V_m= -21.5\ mV$ (resp. $g_{kCa}= -52.87019197\ s^{-1}$) to $V_m=-48.3\ mV$ (resp. $g_{K,Ca}= 32.92\ s^{-1}$) to predict the locally active region in memristive Chay model of excitable membrane. Observe from Fig. 18(a), the real part of the admittance of the frequency response $ReY(i\omega; V_m(Q))>0$ at $V_m= -21.5\ mV$ (resp. $g_{kCa}= -52.87019197\ s^{-1}$), thereby confirming that the it is locally passive. However, when $V_m<-21.5mV$, our in depth simulation in Fig. 18(b) shows that $ReY(i\omega;V_m(Q))=0$ at $V_m= -22.1378795486045\ mV$ (resp. $g_{K,Ca}= -45.5241957133932\ s^{-1}$) and Fig. 18(c) and Fig. 18(d) show that $ReY(i\omega; V_m(Q))< 0$ at $V_m= -24.5$ (resp. $g_{K,Ca}= -23.006\ s^{-1}$) and $V_m= -48.1\ mV$ (resp. $g_{K,Ca}= 31.27\ s^{-1}$) respectively for some frequency ω, confirming that it is locally active. Our simulations in Fig. 18(e) shows that a further decrease in the DC equilibrium voltage at $V_m$= -48.2459999459569 mV( resp. $g_{K,Ca}$= 32.4605269226518 $s^{-1}$), the loci is tangential to the ω axis i.e. *Re Y (iω; Vm(Q))*=0. However, when $V_m$<-48.2459999459569 mV, say $V_m$=-48.3mV(resp. $g_{K,Ca}$= 32.9 $s^{-1}$), Fig. 18(f) shows that Re *Y (iω; Vm(Q))*>0 which confirms that it is locally passive. Therefore, the local activity regime which started below $V_m$=-22.1378795486045 mV (resp. $g_{K,Ca}$= -45.5241957133932 $s^{-1}$) exists over the following regime

$$\begin{array}{c} Local\ Activity\ Re\ gime \\ -22.1378795486045\ mV > V_m > -48.2459999459569\ mV \\ -45.5241957133932\ s^{-1} < g_{KCa} < 32.4605269226518\ s^{-1} \end{array}$$

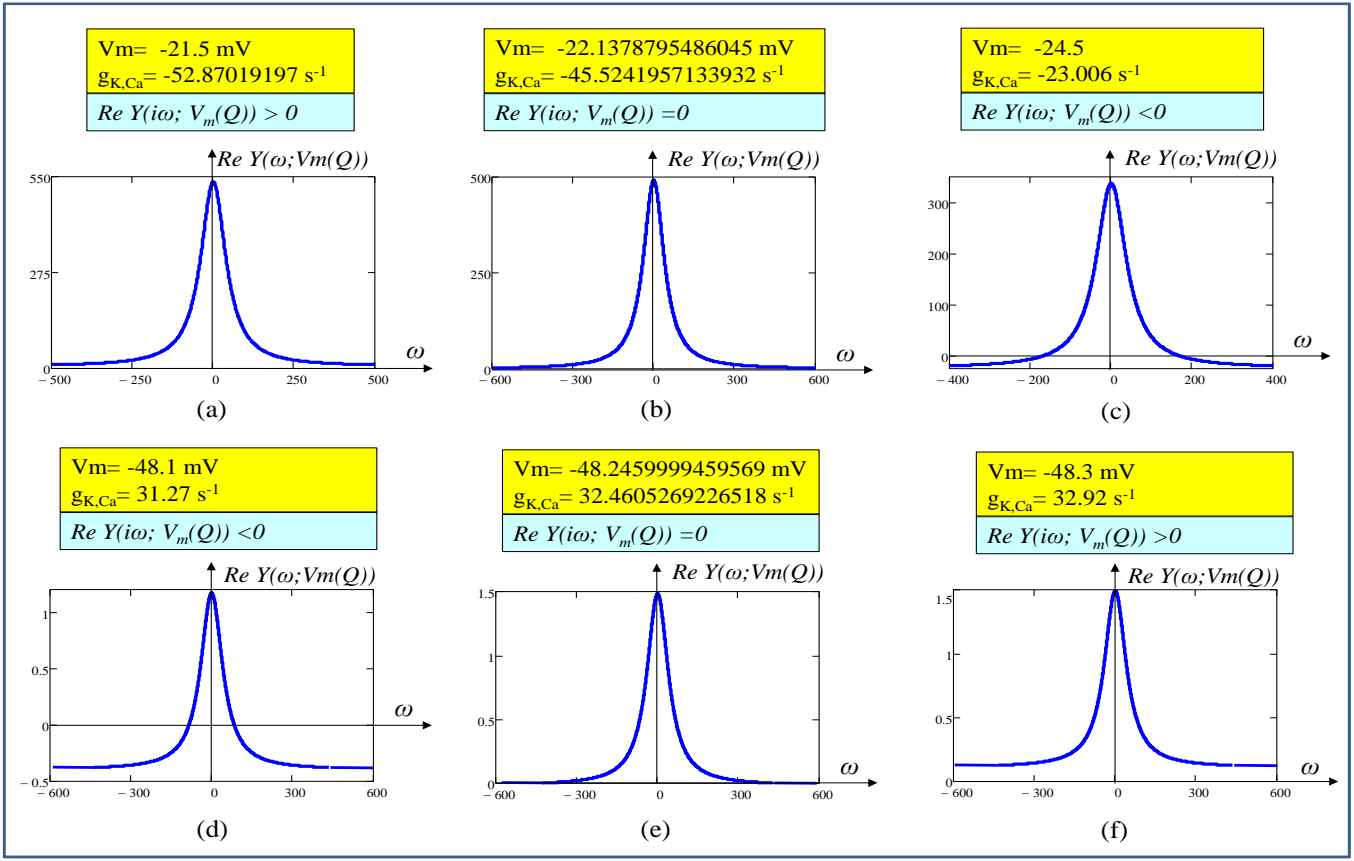

**Fig. 18.** Plot of *Re(iω; Vm(Q))* to illustrate the local activity principle at (a) $V_m$= -21.5 mV (resp. $g_{K,Ca}$= -52.87019197 $s^{-1}$) (b) $V_m$= -22.1378795486045 mV (resp. $g_{K,Ca}$= -45.5241957133932 $s^{-1}$) (c) $V_m$= -24.5 (resp. $g_{K,Ca}$= -23.006 $s^{-1}$) (d) $V_m$= -48.1 mV (resp. $g_{K,Ca}$= 31.27 $s^{-1}$) (e) $V_m$= -48.2459999459569 mV( resp. $g_{K,Ca}$= 32.4605269226518 $s^{-1}$) (f) $V_m$=-48.3mV(resp. $g_{K,Ca}$= 32.9 $s^{-1}$), respectively.

### 5.2 Edge of chaos regime

Edge of chaos is a tiny subset of the locally-active domain where the zeros of admittance function *Y(s; Vm(Q))*(equivalent to the Eigen values of Jacobean matrix) lie in the open left-half plane, i.e. Re($s_p$)<0( Eigen values λi<0) as well as *ReY(iω; Vm(Q))*<0. Fig. 17(a) and Fig. 17(b) show that the real part of the *Eigen values* vanishes at $V_m$= -26.75527972 mV(resp. $g_{K,Ca}$= -7.79022731 $s^{-1}$) with pair of complex Eigen values $\lambda_{2,3}$= ± 97.171320482i . It follows from the edge of chaos theorem that the corresponding equilibrium point is no longer *asymptotically* stable, and becomes unstable thereafter confirming the *1st edge of chaos* regime over the following small interval:

$$\begin{array}{c} Edge\ of\ Chaos\ domain\ 1 \\ -22.1378795486045\ mV > V_m > -26.75527972\ mV \\ -45.5241957133932\ s^{-1} < g_{KCa} < -7.79022731\ s^{-1} \end{array}$$

Observe from Fig. 17(c) that the real Eigen values of the Jacobean Matrix vanishes at $\lambda_{2,3}$= ±1.06i at DC equilibrium voltage $V_m$=-47.5332788572 mV(resp. $g_{K,Ca}$=27.25111606 $s^{-1}$). It follows that the corresponding equilibrium point $V_m(Q)$ is no longer *asymptotically* stable above this equilibrium point, therefore confirming the existence of a *2nd edge of chaos* regime over the following interval:

$$\begin{array}{c} Edge\ of\ Chaos\ Domain\ 2 : \\ -48.2459999459569\ mV < V_m < -47.5332788572\ mV \\ 32.4605269226518\ s^{-1} > g_{KCa} > 27.25111606\ s^{-1} \end{array}$$

## 5.3 Hopf-Bifurcation

Hopf-bifurcation is a locally bifurcation phenomenon in which an equilibrium point changes its stability as the parameter of the nonlinear system changes under certain conditions. There are two types of Hopf-bifurcations namely: *super-critical* Hopf bifurcation and *sub-critical* Hopf bifurcation. An unstable equilibrium point surrounded by a stable limit cycle results to a super-critical Hopf bifurcation and unstable limit cycle surrounded by a stable equilibrium point results to a sub-critical Hopf bifurcation. Our careful simulation at Hopf-bifurcation point 1 at $V_m$= -26.75527972 mV(resp. $g_{K,Ca}$= -7.79022731 $s^{-1}$ ) [8] shows that $g_{KCa}$ chosen within very small edge of chaos domain 1, where the real part of the Eigen values are negative, the result converges to DC equilibrium for any initial conditions. Likewise, $g_{KCa}$ selected within the bifurcation point 1, where the real part of Eigen values are positive, the result converges to a stable limit cycle. Therefore, it follows from the bifurcation theory that bifurcation point 1 is a super-critical Hopf bifurcation. Fig. 19(a) and Fig. 19(b) show the numerical simulation at $g_{K,Ca}$=-7.8 $s^{-1}$ and $g_{K,Ca}$=-7.78 $s^{-1}$ respectively. Observe, from Fig. 19(a) and Fig. 19(b) that, $g_{K,Ca}$=-7.8 $s^{-1}$ lying within the tiny subset of edge of chaos domain 1 converge to DC equilibrium and $g_{K,Ca}$=-7.78 $s^{-1}$ which lies in open half right-plane converge to a stable limit cycle respectively, confirming the bifurcation point 1 is a super-critical Hopf bifurcation.

Similarly, our careful examination predicts a small unstable spike trains when a $g_{KCa}$ is chosen within a very small edge of chaos domain 2 where the Eigen values of the real part are negative and beyond the bifurcation point 2, $V_m$= -47.5332788572 mV(resp. $g_{K,Ca}$=27.25111606 $s^{-1}$). A simple deviation in the initial condition converges to a stable DC equilibrium which confirms the bifurcation point 2 is a sub-critical Hopf bifurcation[9]. The possible scenario of the sub-critical Hopf bifurcation is illustrated in Fig. 20. The upper waveform in Fig. 20(a) shows the membrane potential $V_m$ converges to spikes at the given initial condition indicated in the figure when $g_{KCa}$=27.25345 $s^{-1}$ which lies within edge of chaos domain 2. However, a small perturbation in the initial condition from $V_m(0)$=-47 mV to $V_m(0)$=-48 mV results to converge a DC equilibrium as shown in Fig. 20(b) confirming the bifurcation point 2 is a sub-critical Hopf bifurcation. Fig. 20(c) shows when $g_{KCa}$ =27.250 $s^{-1}$ chosen close and inside the bifurcation point 2, where the real part of the Eigen values are positive, the transient waveform converges to spikes as predicted by Hopf bifurcation theorem.

---

[8] The Hopf bifurcation point 1 occur at the negative of $g_{K,Ca}$ (i.e $g_{K,Ca}$ = -7.79022731 $s^{-1}$) . Negative conductance has no physical significance.

[9] Generally a subcritical Hopf bifurcation predicts a small unstable sinusoidal limit cycle surrounded by a stable DC equilibrium point. In this study, the parameter $g_{KCa}$ , which lies very close and inside the bifurcation point 2 where the complex conjugate of the real part of Eigen values are positive, the transient waveform converges to spikes. Therefore, it is also predicted when $g_{KCa}$ is chosen near and beyond the bifurcation point 2, and within the tiny subset of the edge of chaos domain 2, where the complex conjugate of the real part of the Eigen values are negative, an unstable spike train is surrounded by stable DC equilibrium. We also caution the readers that the above scenario is predicted in this study. However, the scenario might be different in different studies depending on the characteristics of the differential equations.

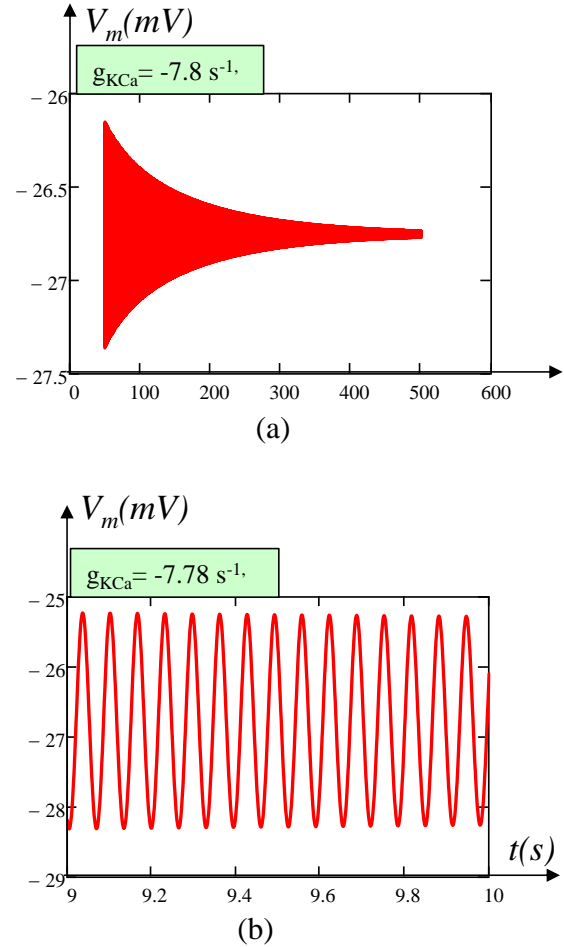

**Fig. 19.** Numerical simulations to confirm the super-critical Hopf bifurcation at bifurcation point 1. Plot of membrane potential $V_m$ at (a) $g_{K,Ca}$=-7.8 $s^{-1}$ which lies inside the tiny subset of edge of chaos domain 1 and beyond bifurcation point 1 converges to the DC equilibrium, (b) and $g_{KCa}$=-7.78 $s^{-1}$, which lies close and open half right-plane of the bifurcation point 1 converges to the stable limit cycle.

Table 7 illustrates the computation of the potassium ion-channel activation $n$, calcium concentration $Ca$ and Eigen values ($\lambda_1$, $\lambda_2$ and $\lambda_3$) as a function of the calcium sensitive potassium conductance $g_{K,Ca}$(resp. membrane potential $V_m$) at the DC equilibrium point $Q$. Observe from Table 7 and Fig. 17(a)-Fig. 17(c) that two Hopf bifurcations points 1 and 2 occur at $V_m$= -26.75527972 mV(resp. $g_{K,Ca}$= -7.79022731 $s^{-1}$ ) and $V_m$= -47.5332788572 mV(resp. $g_{K,Ca}$=27.25111606 $s^{-1}$) respectively, where the real parts of the Eigen values are zero at these equilibria. As $g_{Kca}$ decreases (resp. $V_m$ increases) from

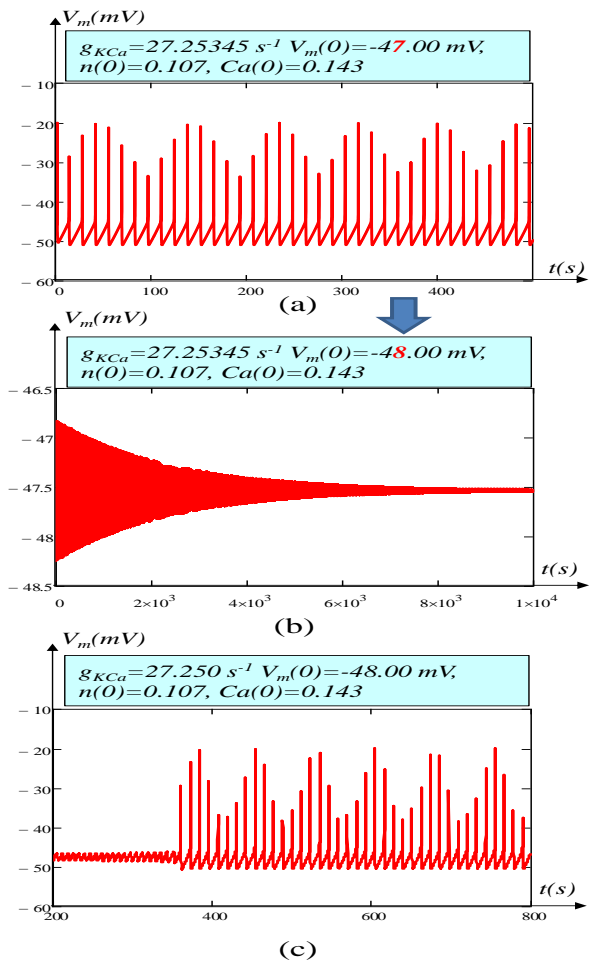

the second regime of negative Eigen values over the following interval:

*Negative real Eigen Values Regime 2:*
$+\infty < V_m < -47.5332788572$ mV
$+\infty > g_{KCa} > 27.25111606$ $s^{-1}$

Observe from Table 7 and Fig. 17(a)-Fig. 17(c) that the complex conjugates and real part of the Eigen values between the two bifurcation on $I_m(\lambda)$ versus $Re(\lambda)$ plane lie on the right hand plane are positive, confirming an unstable DC equilibrium regime and mechanism to give the birth of oscillation, bursting, spikes and chaos in memristive Chay model over the following interval[10]:

*Unstable ( periodic, bursting, chaos, spikes ) Regime :*
$-47.5332788572$ mV $< V_m < -26.75527972$ mV
$27.25111606$ $s^{-1} > g_{KCa} > -7.79022731$ $s^{-1}$

In order to confirm the memristive Chay model converges to a stable DC equilibrium when the real part of the complex conjugate of zeros are negative and little far from the two Hopf bifurcation points, the simulation results at $g_{K,Ca}= -8$ $s^{-1}$ and $g_{K,Ca}=27.3$ $s^{-1}$ are shown in Fig. 21(a) and Fig. 21(b) respectively. Observe from Fig. 21(a) and Fig. 22(b) that the membrane potential $V_m$ converges to a stable DC equilibrium. Similarly, when $g_{K,Ca}= -7.7$ $s^{-1}$ and $g_{K,Ca}= 27.2$ $s^{-1}$, which lie inside the two bifurcation points and the corresponding Eigen values are located in the open right hand in $Im(\lambda_i)$ vs. $(Re\lambda_i)$ plane converge to the unstable DC equilibrium (oscillation) according to the Hopf bifurcation theorem. The corresponding oscillations observed at $g_{K,Ca}= -7.7$ $s^{-1}$ and $g_{K,Ca}= 27.2$ $s^{-1}$ are shown in Fig. 21(c) and Fig. 21(d) respectively.

Fig. 22(a)-Fig. 22(f) show the different pattern of oscillations when the conductance $g_{KCa}$ of calcium sensitive potassium ion channel memristor is varied from $10$ $s^{-1}$ to $11.5$ $s^{-1}$. Fig. 22(a) shows that the excitable membrane cell has a stable limit cycle oscillation with period one at $g_{K,Ca}=10$ $s^{-1}$. As the parameter $g_{K,Ca}$ increases to 10.7 $s^{-1}$, 10.75 $s^{-1}$ and 10.77 $s^{-1}$ the cell fires period two, four and eight as shown in Fig. 22(b), Fig. 22(c) and Fig. 22(d) respectively. The change in the period doubling is more apparent in calcium concentration *(Ca) versus time* and, $V_m$ *versus Ca* as shown in the bottom of Fig. 22(b), Fig. 22(c) and Fig. 22(d) respectively. Fig. 22(e) shows the waveform of the memrisive Chay model confirming the existence of aperiodic oscillation (chaos) at $g_{K,Ca}=11$ $s^{-1}$. The firing of aperiodic oscillations from cell can be clearly seen from the plot of the *Ca versus time* and $V_m$ *versus Ca* in Fig. 22 (e). A further increase in $g_{K,Ca}$ to 11.5 $s^{-1}$ gives rise to the firing of the cell from aperiodic to rhythmic bursting as shown in fig. 22(f).

**Fig. 20.** Numerical simulations to confirm the sub-critical Hopf bifurcation at bifurcation point 2. (a) Plot of membrane potential $V_m$ which converges to spikes at the initial conditions $V_m(0)=-47$ mV, n(0)=0.107 and Ca(0)=0.143 when $g_{K,Ca}=-27.25345$ $s^{-1}$ chosen inside the tiny subset of edge of chaos domain 2 and, near and beyond the bifurcation point 2. (b) Plot of the membrane potential for the same parameters of Fig. 20(a) except the initial condition $V_m(0)$ is changed to $V_m(0) =-48$ mV. The transient waveform converges to DC equilibrium. (c) Plot of membrane potential converging to spikes as predicted by Hopf bifurcation theorem when $g_{K,Ca}=27.250$ $s^{-1}$ is chosen inside the bifurcation point (open half-right pane).

the Hopf bifurcation point 1, the Eigen values migrate to the left hand side confirming that the real parts of the complex conjugate of the Eigen values are no longer positive and thereby confirming the stable and first regime of real negative Eigen values over following interval.

*Negative real Eigen Values Regime 1:*
$-\infty > V_m > -26.75527972$ mV
$-\infty < g_{KCa} < -7.79022731$ $s^{-1}$

Similarly, as $g_{KCa}$ increases(resp. $V_m$ decreases) from the second bifurcation points, the Eigen values migrated to the left hand side confirming that the real part of the complex conjugates of Eigen values are negative, there by confirming

---

[10] The interval where the unstable spike train occurs by stable equilibrium point near the sub-critical Hopf bifurcation is not included in the unstable (periodic, chaos, bursting, spikes) regime.

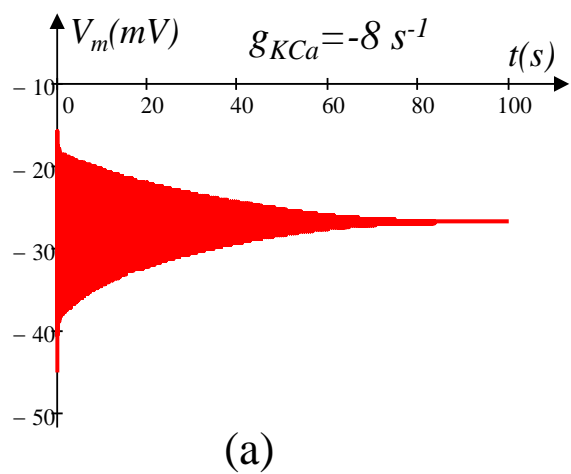
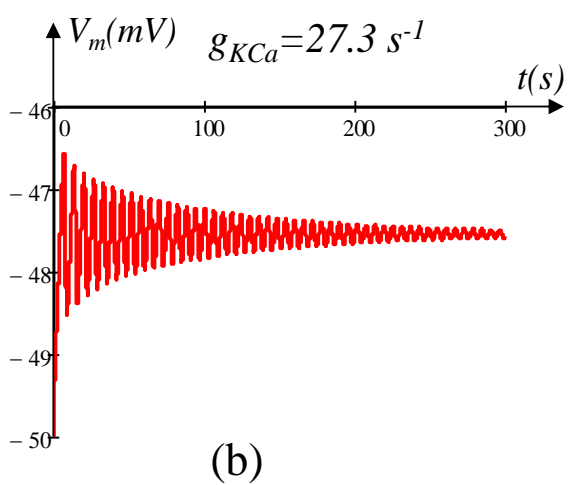
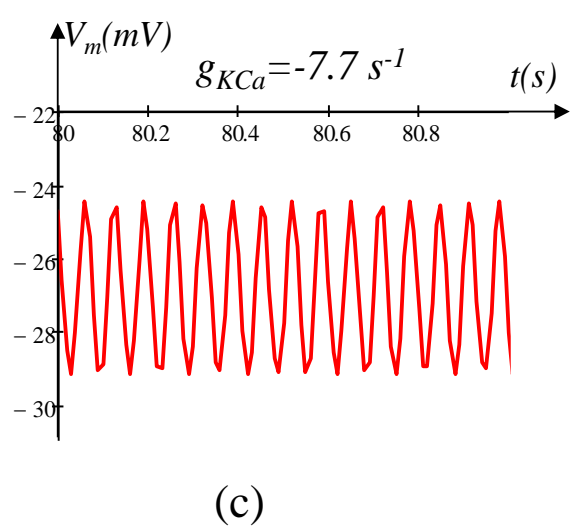
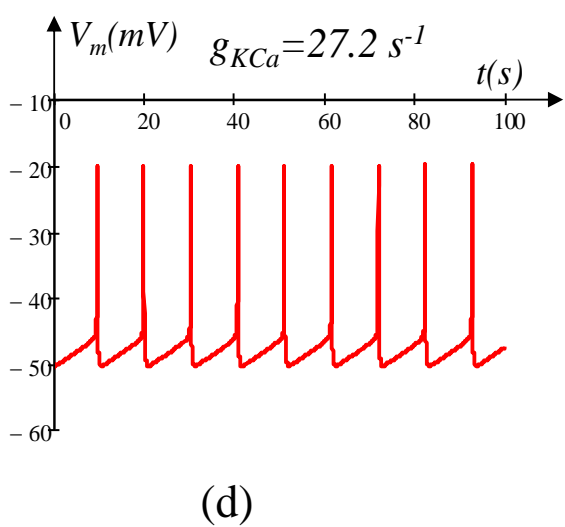

**Fig. 21.** Plot of membrane potential beyond and inside the bifurcations points at (a) $g_{K,Ca}=-8\ s^{-1}$ (b) $g_{K,Ca}=27.3\ s^{-1}$ (c) $g_{K,Ca}=-7.7\ s^{-1}$ and (d) $g_{K,Ca}=27.1\ s^{-1}$. Fig. 21(a) and Fig. 21(b) illustrate the convergence of membrane potential to DC equilibria when $g_{K,Ca}=-8\ s^{-1}$ and $g_{K,Ca}=27.3\ s^{-1}$ are chosen beyond the bifurcation points(open- half left plane). Fig. 21(c) and Fig. 21(d) show the convergence of membrane potential to stable limit cycle and spikes when the parameter $g_{K,Ca}=-7.7\ s^{-1}$ and $27.2\ s^{-1}$ lie inside the two bifurcation points(open half-right plane).

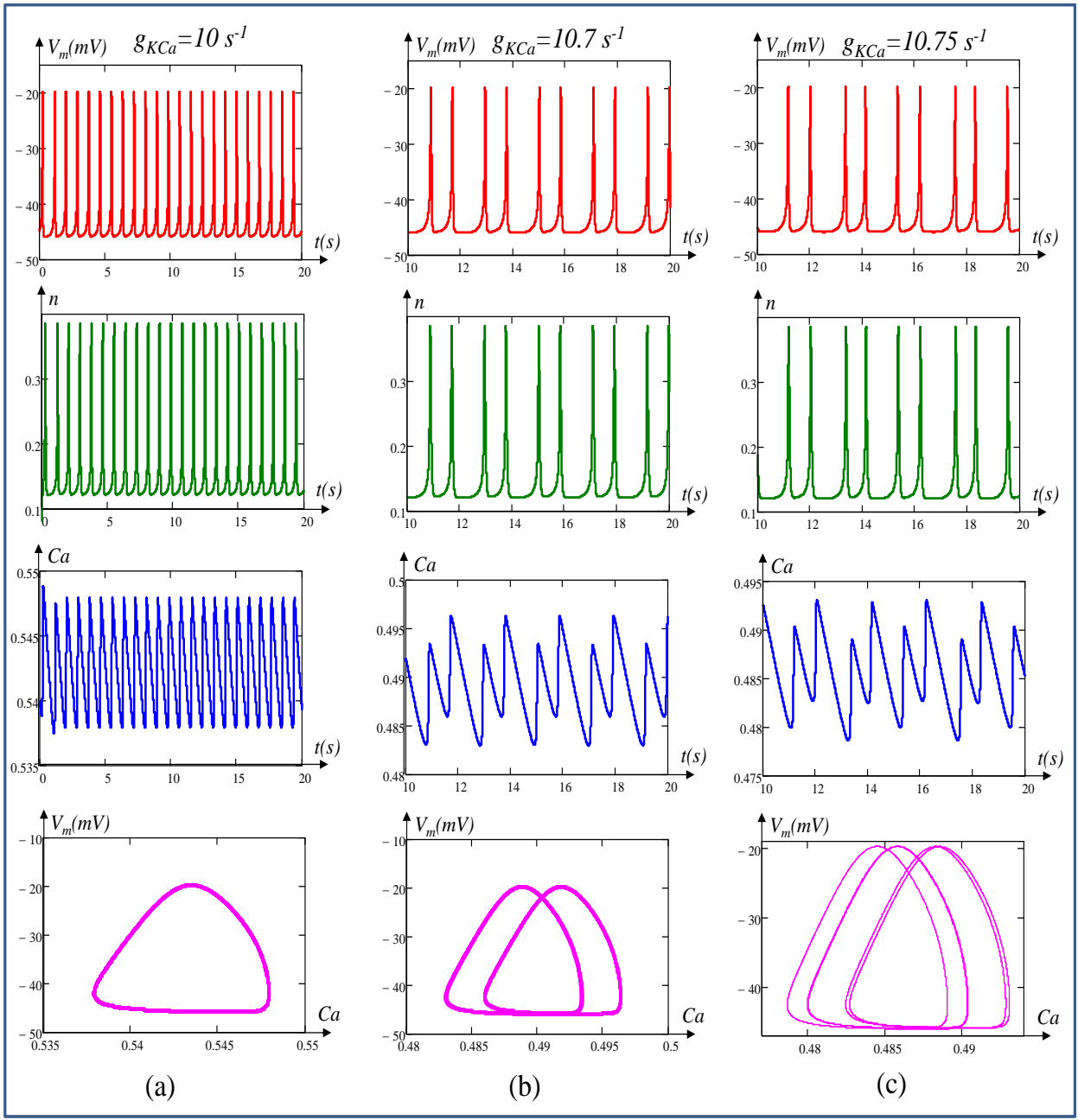

**Fig. 22.** Different pattern of oscillations when $g_{KCa}$ varied from $10\ s^{-1}$ to $11.5\ s^{-1}$. (a) Period-1 oscillation at $g_{K,Ca}=10\ s^{-1}$ (b) Period-2 oscillation at $g_{K,Ca}=10.7\ s^{-1}$ (c) Period-4 oscillation at $g_{K,Ca}=10.75\ s^{-1}$ (d) Period-8 oscillation at $g_{K,Ca}=10.77\ s^{-1}$ (e) Aperiodic (chaotic) oscillation at $g_{K,Ca}=11\ s^{-1}$ (f) Bursting at $g_{K,Ca}=11.5\ s^{-1}$. The simulations were performed at the initial conditions $V_m(0)=-50mV$, $n(0)=0.1$ and $Ca(0)=0.48$.

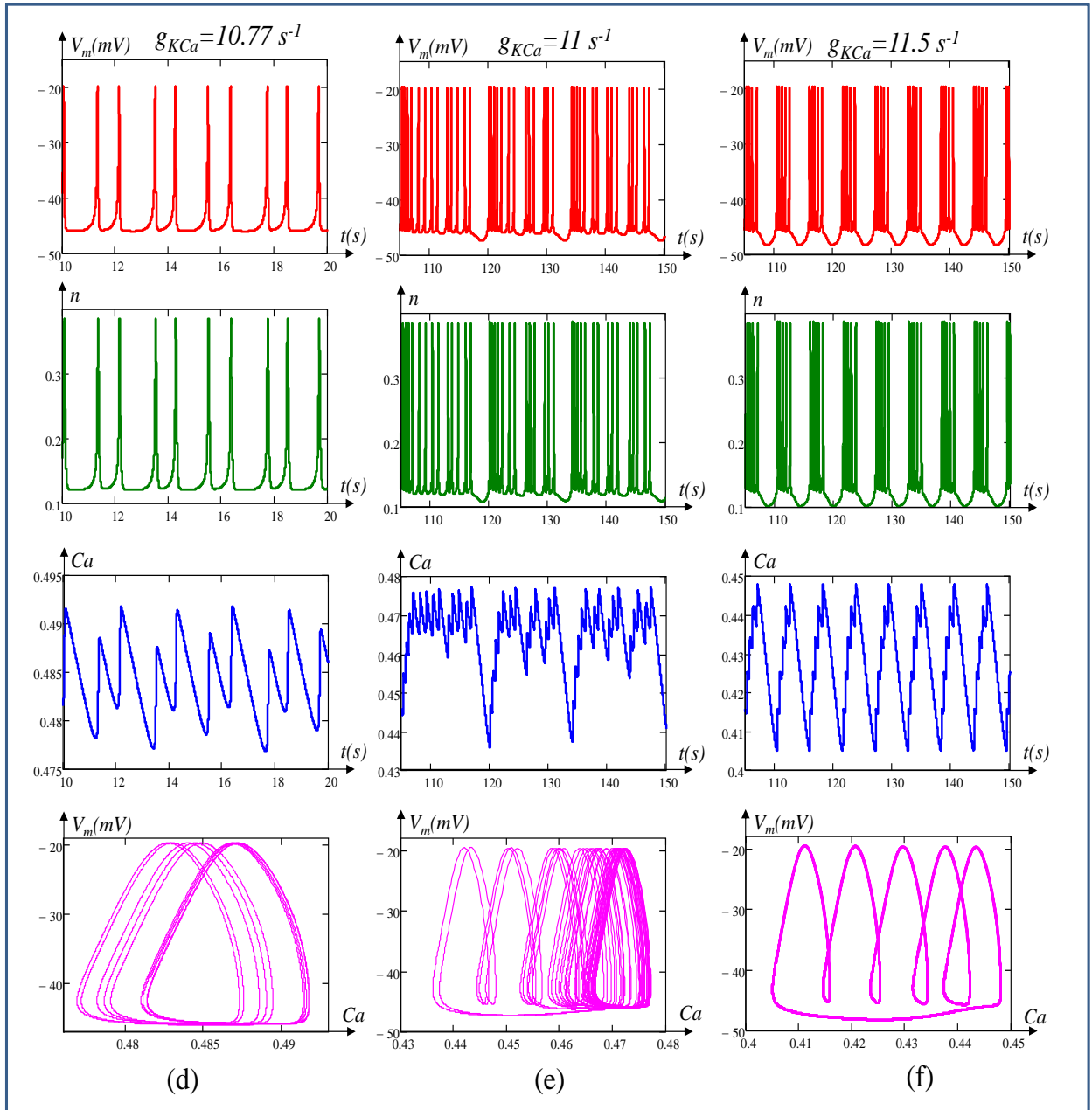

**Fig. 22**(continued)

**Table 7:** Computation of the potassium ion-channel activation *n*, calcium concentration *Ca* and Eigen values ($\lambda_1$, $\lambda_2$ and $\lambda_3$) as a function of the calcium sensitive potassium conductance $g_{K,Ca}$ (resp. membrane potential $V_m$).

| S.N | $V_m$(Vm) | $g_{KCa}$ (s$^{-1}$) | n | Ca | $\lambda_1$ | $\lambda_2$ | $\lambda_3$ |
|---|---|---|---|---|---|---|---|
| 1. | -50.00 | 54.068 | 0.089 | 0.072 | -39.593 | -0.327 | -4.531 |
| 2. | -49.5 | 46.247 | 0.093 | 0.083 | -39.352 | -0.389 | -3.613 |
| 3. | -49.00 | 39.889 | 0.096 | 0.096 | -39.097 | -0.52 | -2.564 |
| 4. | -48.5 | 34.712 | 0.1 | 0.11 | -38.829 | -1.05+0.397i | -1.05-0.397i |
| 5. | -48.2459999459569 | *32.4605269226518* | 0.102 | 0.118 | -38.687 | -0.788+0.778i | -0.788-0.778i |
| 6. | -48.00 | 30.49 | 0.104 | 0.126 | -38.545 | -0.524+0.957i | -0.524-0.957i |
| 7. | -47.5332788572 | 27.25111606 s$^{-1}$ | 0.107 | 0.143 | -38.263 | 0+1.061i | 0-1.061i |
| 8. | -47.00 | 24.225 | 0.112 | 0.165 | -37.921 | 0.639+0.803i | 0.639-0.803i |
| 9. | -46.71 | 22.832 | 0.114 | 0.178 | -37.725 | 1.005+0.058i | 1.005-0.058i |
| 10. | -46.7087457175 | 22.8259856196 | 0.114 | 0.178 | -37.724 | 1.007 | 1.007 |
| 11. | -46.00 | 20.035 | 0.12 | 0.213 | -37.211 | 3.663 | 0.251 |
| 12. | -45.00 | 17.237 | 0.129 | 0.272 | -36.396 | 6.745 | 0.117 |
| 13. | -40.00 | 12.766 | 0.181 | 0.792 | -29.679 | 24.533 | 0.006463 |
| 14. | -34.2426602517 | 11.713175239 | 0.255 | 1.971 | -0.286 | 10.77 | 10.77 |
| 15 | -34.2426602516 | 11.7131752389 | 0.255 | 1.971 | -0.286 | 10.7701+0.0002i | 10.7701-0.0002i |
| 16. | -30 | 5.285 | 0.318 | 3.119 | -0.052 | 9.833+60.283i | 9.833-60.283i |
| 17. | -26.75527972 | -7.79022731 | 0.368 | 3.948 | -0.049 | 0+97.171i | 0-97.171i |
| 18. | -26.7435186728 | -7.8552277404 | 0.369 | 3.95 | -0.04869 | -0.049+97.306i | -0.049-97.306i |
| 19. | -26.7435186727 | --7.8552277409 | 0.369 | 3.95 | -0.04869-97.30591i | -0.0487 | -0.0487+97.3059i |
| 20. | -26.00 | -12.258 | 0.38 | 4.115 | -3.283-105.814i | -0.0485 | -3.283+105.814i |
| 21 | -22.1378795486045 | -45.5241957133932 | 0.442 | 4.737 | -23.21-149.661i | -0.0487 | -23.21+149.661i |
| 22 | -21.00 | -2075.547 | 0.863 | 0.836 | -66.556-438.256i | -0.133 | -66.556+438.256i |

## 6. Concluding Remarks

This paper presented the comprehensive and quantitative analysis of the biological excitable cell based on the Chay neuron model. We proved from the memristive theory that the *voltage-sensitive potassium ion-channel* and *calcium-sensitive potassium ion-channel* in excitable cells are in fact time invariant first-order generic memristors, and *voltage-sensitive mixed ion-channel* is in fact nonlinear resistor.

We also presented in-depth analysis to derive the small signal model, admittance function, pole-zero diagram, frequency response of admittance functions, Nyquist plot and etc., at the DC equilibrium point *Q*. We proved from local activity, edge of chaos theorem and extensive simulations that the local activity regime in memrisitve Chay model exists over the range *-45.5241957133932 s$^{-1}$< $g_{K,Ca}$ <32.4605269226518 s$^{-1}$*, and *edge of chaos regime domain 1* and *domain 2* exist over the range *-45.5241957133932 s$^{-1}$< $g_{K,Ca}$ < -7.79022731 s$^{-1}$* and *32.4605269226518 s$^{-1}$> $g_{K,Ca}$ >27.25111606 s$^{-1}$* respectively. As predicted by Hopf bifurcation theorem, the periodic, periodic-doubling, chaotic, bursting and spike regime are found to exist between two bifurcation points over the range *-7.79022731 s$^{-1}$< $g_{K,Ca}$ <27.25111606 s$^{-1}$*. In accordance with bifurcation theorem, numerical simulations showed that the complex conjugates of Eigen values coincide in purely imaginary axis at *±97.171320482i* and *±1.06i* are *super-critical* and *sub-critical* Hopf bifurcation points, respectively. It is also found that the change in parameter $g_{KCa}$ in excitable cells far from the bifurcation points no longer hold the bifurcation theorem, as it crosses the imaginary axis from right to left confirming the real part of the Eigenvalues are no longer positive and converge to a DC equilibrium.